\documentclass[journal,twoside,twocolumn,a4paper]{IEEEtran}
\usepackage{balance}
\usepackage{amsmath,amssymb,dsfont,color}
\usepackage{graphicx,wrapfig,multicol,multirow}
\usepackage{psfrag,float,cite}

\newcommand{\xpol}{{\mathsf{x}}}
\newcommand{\ypol}{{\mathsf{y}}}
\newcommand{\ppol}{{\mathsf{p}}}

\newcommand{\tnr}[1]{{\textnormal{#1}}}
\newcommand{\tr}[1]{\mathrm{#1}}
\newcommand{\ld}{\ldots}
\newcommand{\ms}[1]{\mathds{#1}}

\newcommand{\bC}{\boldsymbol{C}}\newcommand{\bc}{\boldsymbol{c}}
\newcommand{\bB}{\boldsymbol{B}}\newcommand{\bb}{\boldsymbol{b}}
\newcommand{\bX}{\boldsymbol{X}}\newcommand{\bx}{\boldsymbol{x}}
\newcommand{\bY}{\boldsymbol{Y}}\newcommand{\by}{\boldsymbol{y}}

\newcommand{\bL}{\boldsymbol{L}}

\newcommand{\mcC}{\mathcal{C}}

\newcommand{\mcX}{\mathcal{X}}

\newcommand{\mcXkb}{\mathcal{X}_{k}^{b}}
\newcommand{\mcXko}{\mathcal{X}_{k}^{1}}
\newcommand{\mcXkz}{\mathcal{X}_{k}^{0}}

\newcommand{\Ns}{n}
\newcommand{\Nspan}{N_{\tnr{s}}}

\newcommand{\Rc}{R_\tr{c}}
\newcommand{\Ex}{\ms{E}}

\newcommand{\un}[1]{\underline{#1}}

\newcommand{\argmax}{\mathop{\mathrm{argmax}}}

\newcommand{\BERin}{\tnr{BER}_{\tnr{pre}}}
\newcommand{\BERout}{\tnr{BER}_{\tnr{post}}}
\newcommand{\set}[1]{\{#1\}}
\newcommand{\figref}[1]{Fig.~\ref{#1}}
\newcommand{\secref}[1]{Sec.~\ref{#1}}

\usepackage[ddmmyyyy]{datetime}
\newcommand{\header}{Preprint, \today, \currenttime}
\title{Replacing the Soft FEC Limit Paradigm in the Design of Optical Communication Systems}
\author{Alex Alvarado, Erik Agrell, Domani\c{c} Lavery, Robert Maher, and Polina Bayvel
\thanks{Research supported by the Engineering and Physical Sciences Research Council (EPSRC) project UNLOC (EP/J017582/1), United Kingdom, and by the Swedish Research Council (VR) under grant no.~2012-5280. This work will be presented in part at the 2015 Optical Fiber Communication Conference (OFC), Los Angeles, CA, Mar. 2015.
}
\thanks{A.~Alvarado, D.~Lavery, R.~Maher, and P.~Bayvel are with the Optical Networks Group, Department of Electronic and Electrical Engineering, University College London, London WC1E~7JE, United Kingdom (email: alex.alvarado@ieee.org).}
\thanks{E.~Agrell is with the Department of Signals and Systems, Chalmers University of Technology, SE-41296 Gothenburg, Sweden.}
}
\begin{document}
\maketitle

\begin{abstract}
The FEC limit paradigm is the prevalent practice for designing optical communication systems to attain a certain bit-error rate (BER) without forward error correction (FEC). This practice assumes that there is an FEC code that will reduce the BER after decoding to the desired level. In this paper, we challenge this practice and show that the concept of a channel-independent FEC limit is invalid for soft-decision bit-wise decoding. It is shown that for low code rates and high order modulation formats, the use of the soft FEC limit paradigm can underestimate the spectral efficiencies by up to 20\%. A better predictor for the BER after decoding is the generalized mutual information, which is shown to give consistent post-FEC BER predictions across different channel conditions and modulation formats. Extensive optical full-field simulations and experiments are carried out in both the linear and nonlinear transmission regimes to confirm the theoretical analysis.
\end{abstract}

\section{Introduction and Motivation}

Forward error correction (FEC) and multilevel modulation formats are key technologies for realizing high spectral efficiencies in optical communications. The combination of FEC and multilevel modulation is known as coded modulation (CM), where FEC is used to recover the sensitivity loss from the nonbinary modulation. While in the past optical communication systems were based on hard-decision (HD) FEC, modern systems use \emph{soft-decision} FEC (\mbox{SD-FEC}).

Current digital coherent receivers are based on powerful digital signal processing (DSP) algorithms, which are used to detect the transmitted bits and to compensate for channel impairments and transceiver imperfections. The optimal DSP should find the most likely \emph{coded sequence}. However, this is hard to realize in practice, and thus, most receivers are implemented suboptimally. In particular, detection and FEC decoding are typically decoupled at the receiver: soft information on the code bits is calculated first, and then, an \mbox{SD-FEC} decoder is used. We refer to this receiver structure as a \emph{bit-wise} (BW) decoder, also known in the literature as a bit-interleaved coded modulation (BICM) receiver\cite{Fabregas08_Book,Alvarado15_Book}, owing its name to the original works \cite{Zehavi92,Caire98}, where a bit-level interleaver was included between the FEC encoder and mapper. In the context of optical communications, BW decoders have been studied, e.g., in \cite{Djordjevic2006_JSQE,Bulow2009,Bulow2011b,Millar14,Hager14a,Alvarado13c}.

An alternative to BW decoders is to use iterative demapping (ID) and decoding, i.e., when the FEC decoder and demapper exchange soft information on the code bits iteratively. This structure is known as BICM-ID and was introduced in \cite{Li97,Brink98,Benedetto98b}. BICM-ID for optical communications has been studied in \cite{Djordjevic2007_JLT,Batshon2009_JLT, Bulow14}, \cite[Sec.~3]{Bulow2011b}, \cite[Sec.~3]{Bulow2011}, \cite[Sec.~4]{Schmalen14}. Due to the inherent simplicity of the (noniterative) BW receiver structure, BICM-ID is not considered in this paper.

For simplicity, researchers working on optical communications typically use offline DSP. In this case, and to meet higher-layer quality of service requirements, the bit-error rate (BER) after FEC decoding---in this paper referred to as \mbox{post-FEC} BER or $\BERout$---should be as low as $10^{-12}$ or $10^{-15}$. Since such low BER values cannot be reliably estimated by Monte-Carlo simulations, the conventional design strategy has been to simulate the system without FEC encoding and decoding, and optimize it for a much higher BER value, the so-called ``FEC limit'' or ``FEC threshold''. The rationale for this approach, which we call the \emph{FEC limit paradigm}, is that a certain BER without coding---here referred to as \mbox{pre-FEC} BER or $\BERin$---supposedly can be reduced to the desired \mbox{post-FEC} BER by previously verified FEC implementations.

The use of FEC limits assumes that the decoder's performance is fully characterized by  $\BERin$, and that different channels with the same $\BERin$ will result in the same $\BERout$ using a given FEC code. Under some assumptions on independent bit errors (which can be achieved by interleaving the code bits), this assumption is justifiable, if the decoder is based on HDs. This is the case for HD-FEC, where the decoder is fed with bits modeled using a binary symmetric channel (BSC). The use of FEC limits, however, has not changed with the adoption of \mbox{SD-FEC} in optical communications, which has made the ``\mbox{SD-FEC} limit'' to become increasingly popular in the optical communications literature.

The application of \mbox{SD-FEC} in optical communications dates back to the pioneering experiments by Puc \emph{et al.} in 1999 \cite{Puc99}, who used a concatenation of a Reed--Solomon code and a convolutional code. Other early studies of SD include block turbo codes \cite{Ait-Sab00, Mizuochi04} and low-density parity-check (LDPC) codes \cite{Vasic02,Vasic03,Djordjevic03}. Another concatenated code suitable for SD decoding was defined for optical submarine systems by the ITU in the G.975.1 standard \cite{ITU-T_G.975.1}. See \cite{Djordjevic09}, \cite{Chang10}, and references therein for further details on \mbox{SD-FEC} in optical communications.

Tables and plots of  $\BERout$ vs.~$\BERin$ were presented in, e.g., \cite{Djordjevic03, Mizuochi04, ITU-T_G.975.1}, under specific choices for the channel, modulation format, and symbol rate. Although this was not suggested when these tables and plots were originally published, the existence of such data has subsequently been adopted to avoid the need for including FEC in system simulations and experiments. This \mbox{SD-FEC} limit paradigm is nowadays very popular in optical communication system design. It has been used for example in the record experiments based on $2048$ quadrature amplitude modulation (QAM) for single-core \cite{Beppu15} and multi-core \cite{Qian13ofc} fibers. It has, however, never been validated to which extent the function $\BERout$ vs.~$\BERin$, determined for one set of system parameters (channel, modulation, symbol rate, etc.), accurately characterizes the same function with other parameters.

Another option to predict the post-FEC BER is to use the \emph{mutual information} (MI) between the input and output of the discrete-time channel. This approach was suggested in \cite{Brueninghaus05,Wan06,Franceschini06} and applied to optical communications in \cite{Leven11}. In \cite{Leven11}, it was shown that the MI is a better metric than the pre-FEC BER in predicting the post-FEC BER, which casts significant doubts on the \mbox{SD-FEC} limit paradigm.

This paper investigates the usage of the \emph{generalized mutual information} (GMI) \cite[Sec.~3]{Fabregas08_Book}, \cite[Sec.~4.3]{Alvarado15_Book} for the same purpose. The GMI, also known as the BICM capacity (or parallel decoding capacity), was introduced in an optical communications context in \cite{Bulow2011b}. The performance of some LDPC codes with four-dimensional constellations over the additive white Gaussian noise (AWGN) channel was evaluated in terms of the GMI in \cite{Alvarado2015_JLT}. With any given LDPC code, an apparent one-to-one mapping was observed between the GMI and the post-FEC BER, regardless of the constellation used. In this paper, which extends the conference version \cite{Alvarado15a}, we investigate this mapping further and show that the GMI is a very accurate post-FEC BER predictor, significantly more accurate than both the pre-FEC BER and the MI, under general conditions\footnote{One of these conditions is that the binary code under consideration is \emph{universal}, i.e., that its performance does not depend on the distribution of the soft information passed to the decoder, but only on the capacity of the channel \cite[Sec.~9.5]{Ryan09_Book}. The universality property of LDPC codes for binary-input memoryless channels was initially discussed in \cite{Jones03, Franceschini06}, later studied in, e.g., \cite{Sason09,Sason11}, and recently for spatially-coupled LDPC codes in \cite{Kudekar13}.}. Consistent results were obtained for the nonlinear optical channel in both linear and nonlinear regimes, for the AWGN channel, for both LDPC codes and turbo codes, for a variety of modulation formats, and also validated by experiments.

This paper is organized as follows. In \secref{Sec:Prel}, the system model is introduced and principles for FEC are reviewed. \secref{Sec:Rates} introduces achievable rates, which are quantified by the MI and GMI. The \mbox{post-FEC} BER prediction is studied in \secref{Prediction}. Conclusions are drawn in \secref{Conclusions}.

\section{Preliminaries}\label{Sec:Prel}

\subsection{Channel and System Model}\label{Sec:Prel.Model}

\begin{figure*}[t]
\begin{center}
\newcommand{\scale}{0.72}
\psfrag{bi}[cc][cc][\scale]{}
\psfrag{CHE1}[cc][cc][\scale]{Binary}
\psfrag{CHE2}[cc][cc][\scale]{Encoder\,\,}
\psfrag{bc1y}[cl][cl][\scale]{$\bC_{1}^{\xpol}$}
\psfrag{bcmy}[cl][cl][\scale]{$\bC_{m}^{\xpol}$}
\psfrag{bc1x}[cl][cl][\scale]{$\bC_{1}^{\ypol}$}
\psfrag{bcmx}[cl][cl][\scale]{$\bC_{m}^{\ypol}$}
\psfrag{bc1hx}[cl][cl][\scale]{$\hat{\bC}{}_{1}^{\ypol}$}
\psfrag{bcmhx}[cl][cl][\scale]{$\hat{\bC}{}_{m}^{\ypol}$}
\psfrag{INT}[cc][cc][\scale]{Interleaver}
\psfrag{MOD1}[cc][cc][\scale]{$M$QAM}
\psfrag{MOD2}[cc][cc][\scale]{Mapper}
\psfrag{ENCODER}[cc][cc][\scale]{Binary-Input Soft-Output Channel}
\psfrag{ENCODER2}[cc][cc][\scale]{Characterized by GMI}
\psfrag{EN0}[Bc][Bc][\scale]{\parbox[b]{10em}{\centering Outer\\FEC Encoder}}
\psfrag{EN}[Bc][Bc][\scale]{\parbox[b]{10em}{\centering Inner\\FEC Encoder}}
\psfrag{SDec}[Bc][Bc][\scale]{\parbox[b]{10em}{\centering Inner\\\mbox{SD-FEC} Decoder}}
\psfrag{HDec}[Bc][Bc][\scale]{\parbox[b]{10em}{\centering Outer\\\mbox{HD-FEC} Decoder}}
\psfrag{MI}[Bc][Bc][\scale]{Mutual Information \eqref{MI.def}}
\psfrag{GMI}[Bc][Bc][\scale]{Generalized Mutual Information \eqref{GMI.def.General}}
\psfrag{BERin}[Bc][Bc][\scale]{Pre-FEC BER \eqref{BERin.def.2}}
\psfrag{Optical}[Bc][Bc][\scale]{Optical Channel}
\psfrag{BSC}[cc][cc][\scale]{Binary Symmetric Channel}
\psfrag{BSC2}[cc][cc][\scale]{Characterized by Crossover Probability $\BERout$}
\psfrag{ddd}[cc][cc][\scale]{$\vdots$}
\psfrag{Xy}[bl][Bl][\scale]{$\bX^{\xpol}$}
\psfrag{Xx}[bl][Bl][\scale]{$\bX^{\ypol}$}
\psfrag{Yy}[br][Br][\scale]{$\bY^{\xpol}$}
\psfrag{Yx}[br][Br][\scale]{$\bY^{\ypol}$}
\psfrag{HD}[cc][cc][\scale]{HD}
\psfrag{LLR1}[cc][cc][\scale]{$M$QAM}
\psfrag{LLR2}[cc][cc][\scale]{Demapper}
\psfrag{lc1y}[cl][cl][\scale]{$\bL_{1}^{\xpol}$}
\psfrag{lcmy}[cl][cl][\scale]{$\bL_{m}^{\xpol}$}
\psfrag{lc1x}[cl][cl][\scale]{$\bL_{1}^{\ypol}$}
\psfrag{lcmx}[cl][cl][\scale]{$\bL_{m}^{\ypol}$}
\psfrag{INT2}[cc][cc][\scale]{Deinterleaver}
\psfrag{DEC1}[cc][cc][\scale]{\mbox{SD-FEC}}
\psfrag{Channel}[cc][cc][\scale]{$f_{\un{\bY}|\un{\bX}}(\un{\by}|\un{\bx})$}
\psfrag{SCC}[cc][cc][\scale]{Staircase\,\,}
\psfrag{DEC2}[cc][cc][\scale]{Decoder\,\,}
\psfrag{BERout}[cc][cc][\scale]{$\BERout$}
\includegraphics[width=.95\textwidth]{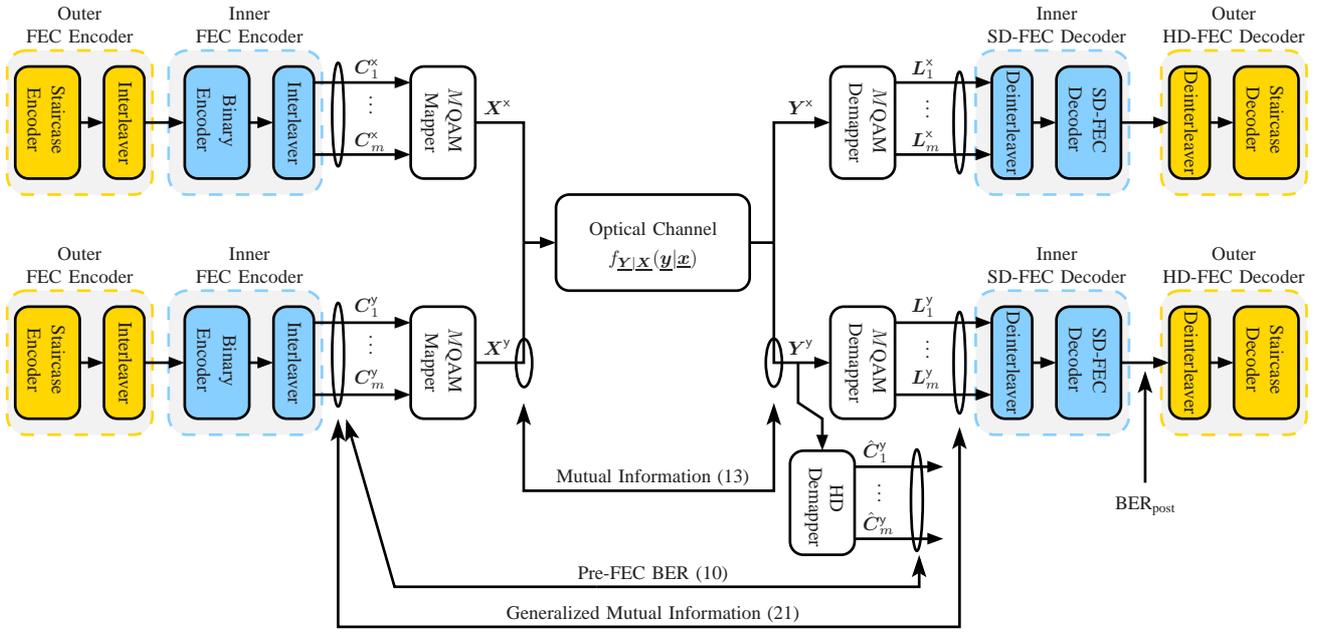}
\end{center}
\caption{Dual-polarization CM transceiver with \mbox{SD-FEC} under consideration. The transmitter for each polarization consists of two cascaded binary FEC encoders followed by an $M$QAM mapper. The receiver is a BW receiver: L-values are calculated by the demapper (ignoring the intersymbol and interpolarization interference), which is followed by an \mbox{SD-FEC} decoder and an \mbox{HD-FEC} decoder.}
\label{model_general_2pol}
\end{figure*}

In this paper, we consider the CM transceiver shown in \figref{model_general_2pol}, which is the common for coherent optical communication systems. Data is transmitted in blocks of $2\Ns$ symbols, where every block represents $\Ns$ time instants in each of the two polarizations. At the transmitter, an outer encoder is serially concatenated with an inner FEC encoder with code rate $\Rc$. The inner encoder generates code bits $\bC_{1}^{\ppol},\ld,\bC_{m}^{\ppol}$, where $\bC_{k}^{\ppol}=[C_{k,1}^{\ppol},C_{k,2}^{\ppol},\ld,C_{k,\Ns}^{\ppol}]$, $k=1,2,\ld,m$ is the bit position and $\ppol\in\set{\xpol,\ypol}$ indicates the polarization.\footnote{Throughout this paper, boldface symbols denote random vectors.} The code bits for each polarization are fed to a memoryless $M$-ary QAM ($M$QAM) mapper with $M=2^{m}$ constellation points $\mcX\triangleq \set{x_{1},x_{2},\ld,x_{M}}$. We consider Gray-mapped square QAM constellations with $M=4,16,64,256$ as well as (non-Gray) $8$QAM from \cite[Fig.~14~(a)]{Ip07}.

The transmitted sequences of complex symbols $\bX^{\ppol}=[X_{1}^{\ppol},X_{2}^{\ppol},\ld,X_{\Ns}^{\ppol}]$ with $X_{l}^{\ppol}\in\mcX$ is modulated using a root-raised-cosine (RRC) pulse with $1\%$ rolloff. The symbols in the two polarizations are combined into the matrix 
\begin{align}
\un{\bX} =
\left[
\begin{matrix}
\bX^{\xpol}\\
\bX^{\ypol}
\end{matrix}
\right]
=
\left[
\begin{matrix}
X_{1}^{\xpol}&X_{2}^{\xpol}&\ld&X_{\Ns}^{\xpol}\\
X_{1}^{\ypol}&X_{2}^{\ypol}&\ld&X_{\Ns}^{\ypol}
\end{matrix}
\right]
\end{align}
and sent through a nonlinear optical channel, whose parameters are summarized in Table~\ref{tab:constants}. We consider 11 dual-polarization wavelength-division multiplexed (WDM) channels of $32~\tnr{Gbaud}$ in a $50$~GHz grid over a single span of single mode fiber (SMF) of length $L$ with zero polarization mode dispersion (PMD). At the receiver, an erbium-doped fiber amplifier (EDFA) with an ideal noise figure of $3$~\tnr{dB} (spontaneous emission factor $n_\tnr{sp}=1$) is used. The digital signal processing (DSP) in the receiver includes electronic chromatic dispersion compensation (EDC) and matched filtering followed by ideal data-aided phase compensation\footnote{In our ideal phase compensation algorithm, the nonlinear phase noise of each received symbol is compensated by multiplying the received symbol by $\exp{(-j\theta_{i})}$ with $i=1,\ldots,M$, where $\theta_{i}$ is the average phase rotation experienced by all the received symbols $Y$ such that $X=x_{i}$.}. Data for the central channel is recorded and represented (for the two polarizations) by the received matrix $\un{\bY}$ of size $2$ by $\Ns$, where $Y_{l}^{\ppol}\in\mathds{C}$ for $l=1,2,\ldots,\Ns$ and $\ppol\in\set{\xpol,\ypol}$.

\begin{table}
	\caption{Summary of system parameters used in WDM simulation.}
	\centering
	\begin{tabular}{ l | l }
\hline
	Parameter		 				& Value \\
\hline 

\hline
		Fiber attenuation			& $0.2~\tnr{dB/km}$ 				\\
		Dispersion parameter		& $17~\tnr{ps}/\tnr{nm}/\tnr{km}$	\\
		Fiber nonlinear coefficient	& $1.2~\tnr{(W km)}^{-1}$ 			\\
		Span length				& $L$~\tnr{km}					\\
		PMD						& $0~\tnr{ps}/\sqrt{\tnr{km}}$		\\
		Symbol rate				& $32~\tnr{Gbaud}$				\\
		EDFA noise figure 			& $3$~\tnr{dB}					\\
		WDM channels				& $11$						\\
		Channel separation			& $50$~\tnr{GHz}				\\
		Pulse shape				& RRC, $1\%$ rolloff				\\
\hline

\hline
	\end{tabular}
	\label{tab:constants}
\end{table}

As shown in \figref{model_general_2pol}, the optical channel is modeled by the channel law $f_{\un{\bY}|\un{\bX}}(\un{\by}|\un{\bx})$.\footnote{Throughout this paper, $f_A(a)$ denotes a probability density function (PDF) and $f_{A|B}(a|b)$ a conditional PDF. Similarly, $P_A(a) \triangleq \Pr\set{A=a}$ denotes a probability mass function (PMF) and $P_{A|B}(a|b) \triangleq \Pr\set{A=a|B=b}$ a conditional PMF.}
This discrete-time model encompasses all the transmitter DSP used after the $M$QAM mapper (i.e., pulse shaping and polarization multiplexing), the physical channel (the fiber and the EDFA), and the receiver DSP.

Even though some residual intersymbol interference usually remains after EDC and the received symbols are affected by interpolarization interference, these effects are typically ignored in current receivers, to reduce complexity. Hence, each symbol in $\un{\bY}$ is decoded separately in both time and polarization. More specifically, for each $l=1,\ld,\Ns$ and $\ppol\in\set{\xpol,\ypol}$, \emph{soft} information on the code bits $C_{1,l}^{\ppol},\ld,C_{m,l}^{\ppol}$ is calculated in the form of \emph{L-values}\footnote{A sign operation on an L-value corresponds an HD. Its magnitude represents the reliability of the HD.}, also known as logarithmic likelihood ratios, as
\begin{align}\label{LLR.def}
L_{k,l}^{\ppol}	& \triangleq \log\frac{f_{Y_{l}^{\ppol}|C_{k,l}^{\ppol}}(y_{l}^{\ppol}|1)}{f_{Y_{l}^{\ppol}|C_{k,l}^{\ppol}}(y_{l}^{\ppol}|0)}\\
\label{LLR.def.diff}
		&= L_{k,l}^{\ppol,\tnr{apo}} - L_{k,l}^{\ppol,\tnr{apri}}
\end{align}
where $k=1,\ldots,m$ and
\begin{align}
\label{LLR.apo.def}
L_{k,l}^{\ppol,\tnr{apo}}&=\log\frac{P_{C_{k,l}^{\ppol}|Y_{l}^{\ppol}}(1|y_{l}^{\ppol})}{P_{C_{k,l}^{\ppol}|Y_{l}^{\ppol}}(0|y_{l}^{\ppol})},\\
\label{LLR.apri.def}
L_{k,l}^{\ppol,\tnr{apri}}&=\log\frac{P_{C_{k,l}^{\ppol}}(1)}{P_{C_{k,l}^{\ppol}}(0)}
\end{align}
are the \emph{a posteriori} and \emph{a priori} L-values, respectively.

A stationary channel model is assumed, and thus, the index $l$ can be dropped. Furthermore, the performance in both polarizations is expected to be identical, so from now on, the notation $({\cdot}){}^{\ppol}$ is also dropped. Using this and the law of total probability in \eqref{LLR.def} gives
\begin{align}\label{LLR.general}
L_{k}	& = \log\frac{\sum_{x\in\mcXko}P_{X|C_{k}}(x|1)f_{Y|X}(y|x)}{\sum_{x\in\mcXkz}P_{X|C_{k}}(x|0)f_{Y|X}(y|x)}
\end{align}
where $\mcXkb\subset\mcX$ is the set of constellation symbols labeled by a bit $b\in\mathbb{B}\triangleq\set{0,1}$ at bit position $k\in\set{1,\ld,m}$.  The L-values calculated by the demapper are then passed to the \mbox{SD-FEC} decoder. The \mbox{SD-FEC} decoder makes a decision on the bits fed into the inner encoder. These bits are then used by the outer \mbox{HD-FEC} decoder, as shown in \figref{model_general_2pol}. 

To alleviate the computational complexity of \eqref{LLR.general}, the well-known max-log approximation \cite{Viterbi98}
\begin{align}\label{LLR.max-log}
L_{k}	& \approx \log\frac{\max_{x\in\mcXko}P_{X|C_{k}}(x|1)f_{Y|X}(y|x)}{\max_{x\in\mcXkz}P_{X|C_{k}}(x|0)f_{Y|X}(y|x)}
\end{align} 
is often used.

\subsection{Pre-FEC BER}\label{Sec:Prel.Model:Pre-FEC}

The lower branch of the receiver in \figref{model_general_2pol} includes an HD demapper which makes an HD on the code bits. We assume that this HD demapper is the optimal memoryless HD demapper in the sense of minimizing the pre-FEC BER. This maximum \emph{a posteriori} (MAP) decision rule is equivalent to making an HD on the \emph{a posteriori} L-values in \eqref{LLR.apo.def}: if $L_{k}^{\tnr{apo}}\geq 0$ then $\hat{C}_{k}=1$, and $\hat{C}_{k}=0$ otherwise.\footnote{This decision rule is slightly better than the standard demapper based on HDs on the symbols followed by a symbol-to-bit mapper (inverting the bit-to-symbol mapping used at the transmitter). However, the differences are noticeable only at very high pre-FEC BER \cite[Sec.~V]{Ivanov13a}.} Formally,
\begin{align}
\label{BERin.def.0}
\BERin 	&\triangleq \frac{1}{m}\sum_{k=1}^{m}\Pr\set{\hat{C}_{k}\neq C_{k}}\\
\label{BERin.def.1}
		&= \frac{1}{m}\sum_{k=1}^{m}\sum_{c\in\mathbb{B}}P_{C_{k}}(c)\Pr\set{\hat{C}_{k}\neq c|C_{k}=c}\\
\label{BERin.def.2}
		&=  \frac{1}{m}\sum_{k=1}^{m}\sum_{c\in\mathbb{B}}P_{C_{k}}(c) \int_{0}^{\infty}f_{L_{k}^{\tnr{apo}}|C_{k}}((-1)^{c}l | c)\,\tnr{d}l.
\end{align}
The pre-FEC BER is a standard performance measure for uncoded systems. As discussed in Sec.~\ref{Sec:Prel.Model:HD-FEC}, pre-FEC BER is a good predictor of post-FEC BER for \mbox{HD-FEC} with ideal interleaving. We will show in Sec.~\ref{Prediction} that the pre-FEC BER is not necessarily a good predictor of post-FEC BER for \mbox{SD-FEC}.

\begin{figure*}[t]
\begin{center}
\newcommand{\scale}{0.75}
\psfrag{bi}[cc][cc][\scale]{}
\psfrag{CHE1}[cc][cc][\scale]{Binary}
\psfrag{CHE2}[cc][cc][\scale]{Encoder\,\,}
\psfrag{bc1}[cc][cc][\scale]{$\,\bC_{1}$}
\psfrag{bcm}[cc][cc][\scale]{$\,\bC_{m}$}
\psfrag{bc1h}[cl][cl][\scale]{$\hat{\bC}_{1}$}
\psfrag{bcmh}[cl][cl][\scale]{$\hat{\bC}_{m}$}
\psfrag{MOD1}[cc][cc][\scale]{$M$QAM}
\psfrag{MOD2}[cc][cc][\scale]{Mapper}
\psfrag{ENCODER}[cc][cc][\scale]{Binary-Input Soft-Output Channel}
\psfrag{ENCODER2}[cc][cc][\scale]{Characterized by GMI}
\psfrag{EN0}[cc][cc][\scale]{\parbox[b]{10em}{\centering Outer\\FEC Encoder}}
\psfrag{EN}[cc][cc][\scale]{\parbox[b]{10em}{\centering Inner\\FEC Encoder}}
\psfrag{SDec}[cc][cc][\scale]{\parbox[b]{10em}{\centering Inner\\\mbox{SD-FEC} Decoder}}
\psfrag{HDec}[cc][cc][\scale]{\parbox[b]{10em}{\centering Outer\\\mbox{HD-FEC} Decoder}}
\psfrag{BSC}[cc][cc][\scale]{Binary Symmetric Channel}
\psfrag{BSC2}[cc][cc][\scale]{Characterized by Crossover Probability $\BERout$}
\psfrag{ddd}[cc][cc][\scale]{$\vdots$}
\psfrag{x}[bc][Bc][\scale]{$\bX$}
\psfrag{y}[bc][Bc][\scale]{$\bY$}
\psfrag{HD}[cc][cc][\scale]{HD}
\psfrag{LLR1}[cc][cc][\scale]{$M$QAM}
\psfrag{LLR2}[cc][cc][\scale]{Demapper}
\psfrag{lc1}[cl][cl][\scale]{$\bL_{1}$}
\psfrag{lcm}[cl][cl][\scale]{$\bL_{m}$}
\psfrag{INT2}[cc][cc][\scale]{Deinterleaver}
\psfrag{DEC1}[cc][cc][\scale]{\mbox{SD-FEC}}
\psfrag{Channel}[cc][cc][\scale]{$f_{\bY|\bX}(\by|\bx)$}
\psfrag{SCC}[cc][cc][\scale]{Staircase\,\,}
\psfrag{DEC2}[cc][cc][\scale]{Decoder\,\,}
\psfrag{BERout}[cc][cc][\scale]{$\BERout$}
\psfrag{a}[Bc][Bc][\scale]{(a) Interface for \mbox{SD-FEC}}
\psfrag{b}[Bc][Bc][\scale]{(b) Interface for HD-FEC}
\includegraphics[width=0.9\textwidth]{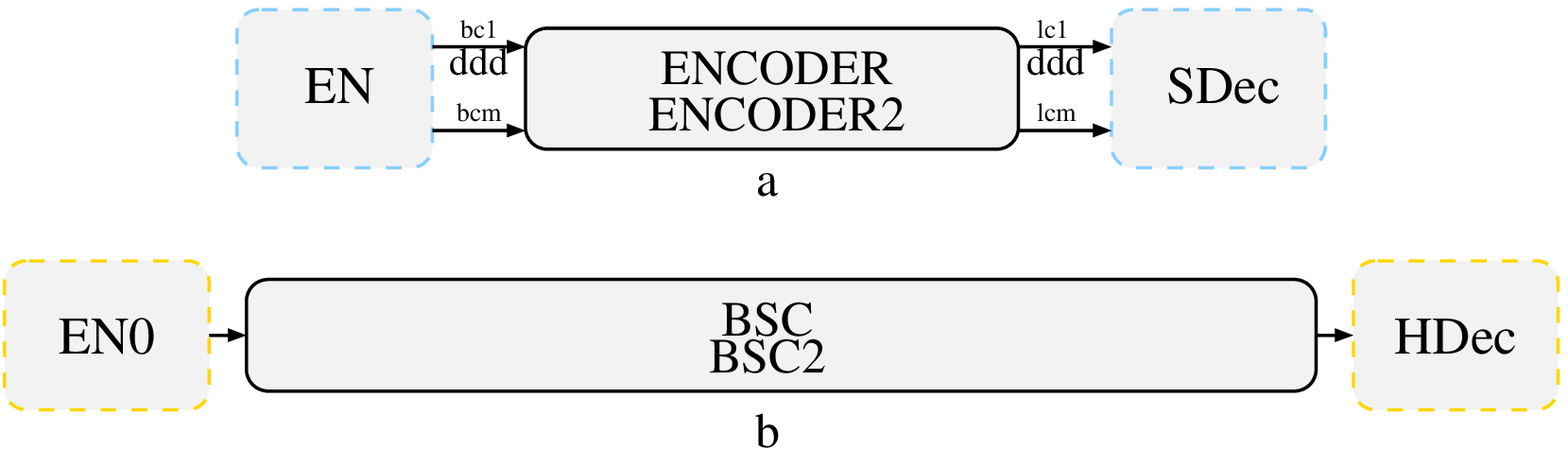}
\end{center}
\caption{Interface for (a) the inner \mbox{SD-FEC} and (b) the outer \mbox{HD-FEC} in \figref{model_general_2pol} (for one polarization). The BISO channel characterized by the GMI and the BSC by its crossover probability given by the BER after \mbox{SD-FEC} decoding $\BERout$.}
\label{model_BISO}
\end{figure*}

\subsection{\mbox{SD-FEC}}\label{Sec:Prel.Model:SD-FEC}

We consider two families of binary \mbox{SD-FEC}: Turbo codes (TCs) and irregular repeat-accumulate LDPC codes. In both cases, a pseudo-random bit-level interleaver is assumed to be used prior to modulation (see \figref{model_general_2pol}). Without loss of generality, we assume this interleaver to be part of the inner FEC encoder.

The TCs we consider are formed as the parallel concatenation of two identical, eight-state, recursive and systematic convolutional encoders with code rate $1/2$. The generator polynomials are $(1,11/15)_8$ \cite{acikel99} and the two encoders are separated by an internal random interleaver, giving an overall code rate $\Rc=1/3$. Six additional code rates $\Rc\in\set{2/5,1/2,3/5,2/3,3/4,5/6}$ are obtained by cyclically puncturing parity bits using the patterns defined in \cite{acikel99} and \cite{kousa02}, which leads to FEC overheads (OHs) of $\set{200,150,100,66.6,50,33.3,20}\%$. Each transmitted frame consists of $20,000$ information bits. The decoder is based on the max-log-MAP decoding algorithm with ten iterations. The extrinsic L-values exchanged during the iterations are scaled by $0.7$ as suggested in \cite{Vogt00}.

The LDPC codes we consider are those proposed by the second generation satellite digital video broadcasting standard \cite{ETSI_EN_302_307_v121} with code rates $\Rc\in\set{1/3, 2/5, 1/2, 3/5, 3/4, 9/10}$. This leads to OHs of $\set{200, 150, 100, 66.66, 33.3, 11.1}\%$. Each transmitted frame consists of $64,800$ code bits. The decoder uses the message passing algorithm with $50$ iterations and exact L-values.

What the \mbox{SD-FEC} encoder and decoder pair ``sees'' is a binary-input soft-output (BISO) channel. This is shown in \figref{model_BISO}~(a). This BISO channel is sometimes known in the literature as the BICM channel \cite[Fig.~1]{Martinez06} and it has been used to predict the decoder performance via probabilistic models of the L-values \cite[Sec.~5.1]{Alvarado15_Book}. In this paper, we are interested in finding a measure to characterize this BISO channel in order to predict the post-FEC BER across different channels.

\subsection{HD-FEC}\label{Sec:Prel.Model:HD-FEC}

As shown in \figref{model_general_2pol}, the considered transceiver includes an outer encoder to reduce the BER after \mbox{SD-FEC} decoding to $10^{-15}$. For both TCs and LDPC codes, we use the staircase code with $6.25\%$ OH from \cite[Table~I]{Zhang14}. For a BSC, this staircase code guarantees an output BER of $10^{-15}$ for a crossover probability of $4.7\cdot 10^{-3}$. This corresponds to the \mbox{HD-FEC} limit paradigm, which is perfectly justifiable under the BSC assumptions.

To guarantee that the errors introduced by the inner \mbox{SD-FEC} decoder are independent within a frame, we include a bit-level interleaver (see \figref{model_general_2pol}). Under these assumptions, what the \mbox{HD-FEC} encoder and decoder pair ``sees'' is a BSC with crossover probability given by the BER after \mbox{SD-FEC} decoding ($\BERout$). Therefore, the BER after \mbox{HD-FEC} decoding can be assumed to be $10^{-15}$ for $\BERout=4.7\cdot 10^{-3}$. This is shown in \figref{model_BISO}~(b). From now on, we therefore assume the existence of the interleaver and staircase code, and thus, without loss of generality, we focus on a target BER after \mbox{SD-FEC} decoding of $\BERout=4.7\cdot 10^{-3}$.

\section{Achievable Rates}\label{Sec:Rates}

Achievable rates provide an upper bound on the number of bits per symbol that can be reliably transmitted through the channel. In this section we review achievable rates for channels with memory, for optimal decoders, and for BW decoders. These achievable rates will be used in \secref{Prediction} to predict the post-FEC BER.

\subsection{Channels with Memory}\label{Sec:Prel.Rates}

A \emph{coding scheme} consists of a codebook, an encoder, and a decoder. The codebook is the set of codewords that can be transmitted through the channel, where each codeword is a sequence of symbols. The encoder is a one-to-one mapping between the information sequences and codewords. The decoder is a deterministic rule that maps the noisy channel observations onto an information sequence. 

A code rate, in bits per (single-polarization) symbol, is said to be \emph{achievable} at a given block length and for a given average error probability $\varepsilon$ if there exists a coding scheme whose average error probability is below $\varepsilon$. Under certain assumptions on information stability \cite[Sec.~I]{verdu94}, and for any stationary random process $\{X_l\}$ with joint PDF $f_{\un{\bX}}$, an achievable rate for channels with memory (i.e., where symbols are correlated in time and across polarizations) is given by
\begin{align}
\label{MI.Memory}
R^{\tnr{mem}} 	& = \lim_{\Ns\rightarrow\infty} \frac{1}{2\Ns} I(\un{\bX};\un{\bY})
\end{align}
where $I(\un{\bX};\un{\bY})$ is the mutual information defined as
\begin{align}\label{MI.def.memory}
I(\un{\bX};\un{\bY}) \triangleq \Ex_{\un{\bX},\un{\bY}}\left[\log_{2}\frac{f_{\un{\bY}|\un{\bX}}(\un{\bY}|\un{\bX})}{f_{\un{\bY}}(\un{\bY})}\right]
\end{align}
and where $\Ex_{\un{\bX},\un{\bY}}$ denotes the expectation with respect to both $\un{\bX}$ and $\un{\bY}$. The \emph{channel capacity} is the largest achievable rate for which a coding scheme with vanishing error probability exists, in the limit of large block length.


\subsection{Memoryless Receivers}\label{Sec:Prel.MI}

Although the discrete-time optical channel in \secref{Sec:Prel.Model} suffers from intersymbol and interpolarization interference, the standard receiver considered in this paper ignores these effects. In particular, each polarization is considered independently (see \figref{model_general_2pol}), and the soft information on the coded bits is calculated ignoring correlation between symbols in time (see \eqref{LLR.def}). To model these assumptions made by the receiver, the channel is modeled by a conditional PDF $f_{Y|X}(Y|X)$. Therefore, from now on, and without loss of generality, only one polarization is considered. Furthermore, we assume the symbols are independent random variables drawn from a distribution $f_X$. 

An achievable rate for transceivers that ignore intersymbol and interpolarization interference is
\begin{align}\label{MI.def}
I(X;Y) = \Ex_{X,Y}\left[\log_{2}\frac{f_{Y|X}(Y|X)}{f_{Y}(Y)}\right].
\end{align}
where $I(X;Y)$ is the unidimensional version of the MI in \eqref{MI.def.memory}. As expected, $R^{\tnr{mem}}\geq I(X;Y)$\cite[Sec.~III-F]{Essiambre10} and thus, $I(X;Y)$ is a (possibly loose) lower bound on the capacity of the channel with intersymbol and interpolarization interference.

Let $\mcC$ the binary codebook used for transmission and $\un{\bc}$ denote the transmitted codewords as
\begin{align}
\un{\bc} = 
\left[
\begin{matrix}
c_{1,1} & c_{1,2} & \ld & c_{1,\Ns}\\
\vdots	&\vdots	& \ddots & \vdots \\
c_{m,1} & c_{m,2} & \ld & c_{m,\Ns}\\
\end{matrix}
\right].
\end{align}
Furthermore, let $\bB=[B_{1},\ld,B_{m}]$ be a random vector representing the transmitted bits $[c_{1,l},\ld,c_{m,l}]$ at any time instant $l$, which are mapped to the corresponding symbol $X_{l}\in\mcX$ with $l=1,2,\ld,\Ns$. Assuming a memoryless channel, the optimal maximum-likelihood (ML) receiver chooses the transmitted codeword based on an observed sequence $[y_{1},\ld,y_{n}]$ according to the rule
\begin{align}\label{rule.ML}
\un{\bc}^{\tnr{ml}} \triangleq \argmax_{\un{\bc}\in\mcC}\sum_{l=1}^{\Ns}\log f_{Y|\bB}(y_{l}|c_{1,l},\ld,c_{m,l}).
\end{align}
Shannon's channel coding theorem states that reliable transmission with the ML decoder in \eqref{rule.ML} is possible at arbitrarily low error probability if the combined rate of the binary encoder and mapper (in information bit/symbol) is below $I(X;Y)$, i.e., if $\Rc m\leq I(X;Y)$.

For a discrete constellation $\mcX$, the MI in \eqref{MI.def} can be expressed as
\begin{align}
\label{MI.general}
I(X;Y) & = \sum_{x\in\mcX} P_{X}(x) \int_{\mathds{C}}f_{Y|X}(y|x) \log_{2}\frac{f_{Y|X}(y|x)}{f_{Y}(y)}\, \tnr{d}y.
\end{align}
A Monte-Carlo estimate thereof is
\begin{align}
\label{MI.general.MC}
I(X;Y) & \approx \frac{1}{\Ns}\sum_{x\in\mcX}P_{X}(x)\sum_{l=1}^{\Ns} \log_{2}\frac{f_{Y|X}(t^{(l)}|x)}{f_{Y}(t^{(l)})}
\end{align}
where $t^{(l)}$ with $l=1,2,\ld,\Ns$ are independent and identically distributed (i.i.d.) random variables distributed according to the channel law $f_{Y|X}(y|x)$.

\subsection{BW Receivers}\label{Sec:Prel.GMI}

As shown in \figref{model_general_2pol}, the BW decoder considered in this paper splits the decoding process. First, L-values are calculated, and then, a binary SD decoder is used. More precisely, the BW decoder rule is
\begin{align}\label{rule.BW}
\un{\bc}^{\tnr{bw}} \triangleq \argmax_{\un{\bc}\in\mcC}\sum_{l=1}^{\Ns}\log \prod_{k=1}^{m}f_{Y|B_{k}}(y_{l}|c_{k,l}).
\end{align}
The BW decoding rule in \eqref{rule.BW} is not the same as the ML rule in \eqref{rule.ML} and the MI is in general not an achievable rate with a BW decoder.\footnote{An exception is the trivial case of Gray-mapped $4$QAM (i.e., quadrature phase-shift keying, QPSK) with noise added in each quadrature independently. In this case, the detection can be decomposed into two binary phase-shift keying constellations, and thus, ML and BW decoders are identical.}

The BW decoder can be cast into the framework of a mismatched decoder by considering a symbol-wise metric
\begin{align}\label{metric.BW}
q(\bb,y) \triangleq \prod_{k=1}^{m}f_{Y|B_{k}}(y|b_{k}).
\end{align}
Using this mismatched decoding formulation, the BW rule in \eqref{rule.BW} can be expressed as 
\begin{align}\label{rule.BW.2}
\un{\bc}^{\tnr{bw}} = \argmax_{\un{\bc}\in\mcC}\sum_{l=1}^{\Ns}\log q(\bb_{l},y_{l})
\end{align}
where with a slight abuse of notation we use $\bb_{l}=[c_{1,l},\ld,c_{m,l}]^{T}$. Similarly, the ML decoder in \eqref{rule.ML} can be seen as a mismatched decoder with a metric $q(\bb_{l},y_{l})=f_{Y|\bB}(y_{l}|\bb_{l})=f_{Y|X}(y_{l}|x_{l})$ which is ``matched'' to the channel. Using this interpretation, the BW decoder uses metrics matched to the bits $f_{Y|B_{k}}(y|b_{k})$, but not matched to the actual (symbol-wise) channel.

An achievable rate for a BW decoder is the GMI, which represents a bound on the number of bits per symbol that can be reliably transmitted through the channel. The GMI is defined as \cite[eq.~(59)--(60)]{Martinez09} \cite[(4.34)--(4.35)]{Alvarado15_Book}
\begin{align}\label{GMI.def.General}
\tnr{GMI} & \triangleq \max_{s\geq 0} \Ex_{\bB,Y}\left[ \log_{2}\frac{q(\bB,Y)^{s}}{\sum_{\bb\in\mathbb{B}^{m}}P_{\bB}(\bb)q(\bb,Y)^{s}}\right].
\end{align}
For the BW metric in \eqref{metric.BW} and assuming independent bits $B_{1},\ld,B_{m}$, the GMI in \eqref{GMI.def.General} can be expressed as
\begin{align}
\label{GMI.BW.0}
\tnr{GMI} & = \max_{s\geq 0} \sum_{k=1}^{m}\Ex_{B_{k},Y}\left[ \log_{2}\frac{f_{Y|B_{k}}(Y|B_{k})^{s}}{\sum_{b\in\mathbb{B}}P_{B_{k}}(b)f_{Y|B_{k}}(Y|b)^{s}}\right]\\
\label{GMI.BW.1}
		& = \sum_{k=1}^{m}\Ex_{B_{k},Y}\left[ \log_{2}\frac{f_{Y|B_{k}}(Y|B_{k})}{\sum_{b\in\mathbb{B}}P_{B_{k}}(b)f_{Y|B_{k}}(Y|b)}\right]\\
\label{GMI.BW.2}
		& = \sum_{k=1}^{m} I(B_{k};Y)
\end{align}
where \eqref{GMI.BW.0} follows from \cite[Theorem~4.11]{Alvarado15_Book} and \eqref{GMI.BW.1}  from \cite[Corollary~4.12]{Alvarado15_Book} (obtained with $s=1$). The expression in \eqref{GMI.BW.2} follows from the definition of MI in \eqref{MI.def}.

In general, $I(X;Y)\geq\tnr{GMI}$ \cite[Theorem~4.24]{Alvarado15_Book}\footnote{The condition of i.i.d. bits in \cite[Theorem~4.24]{Alvarado15_Book} is not necessary---only independence is needed.}, where the rate penalty $I(X;Y)-\tnr{GMI}$ can be understood as the penalty caused by the use of a suboptimal (BW) decoder. This rate penalty, however, is known to be small for Gray-labeled constellations \cite[Fig.~4]{Caire98}, \cite{Agrell10b,Alvarado12b}, \cite[Sec.~IV]{Alvarado11b}.  

The GMI has not been proven to be the largest achievable rate for the receiver in \figref{model_general_2pol}. For example, a different achievable rate---the so-called LM rate---has been recently studied in \cite[Part~I]{Peng12_Thesis}. Moreover, in the case where unequally likely constellation points are allowed, a new achievable rate has been recently derived in \cite[Theorem~1]{Bocherer14}. Finding the largest achievable rate with a BW decoder remains an open research problem. Despite this cautionary statement, the GMI is known to predict well the performance of CM transceivers based on capacity-approaching \mbox{SD-FEC} decoders. This will be shown in \secref{Prediction}. 

When the L-values are calculated using \eqref{LLR.general}, $I(B_{k};Y)=I(B_{k};L_{k})$ \cite[Theorem~4.21]{Alvarado15_Book}, and thus, the GMI in \eqref{GMI.BW.2} becomes
\begin{align}\label{GMI.LLRs}
\tnr{GMI}	& = \sum_{k=1}^{m} I(B_{k};L_{k})
\end{align}
i.e., the GMI is a sum of \emph{bit-wise} MIs between code bits and L-values. The equality in \eqref{GMI.LLRs} does not hold, however, if the L-values were calculated using the max-log approximation \eqref{LLR.max-log}, or more generally, if the L-values were calculated using  any other approximation. For example, when max-log L-values are considered, it is possible to show that there is a loss in achievable rates. Under certain conditions, this loss can be recovered by adapting the max-log L-values, as shown in \cite{Jalden10,Nguyen11,Szczecinski12a}.

Regardless of the L-value calculation, the GMI in \eqref{GMI.BW.0} can be estimated via Monte-Carlo integration as \cite[Theorem~4.20]{Alvarado15_Book}
\begin{align}\nonumber
\tnr{GMI}	& \approx \sum_{k=1}^{m} H_\text{b}(P_{B_{k}}(0))-\\
\label{GMI.general.MC}
		& \hspace{-.5cm}  \frac{1}{\Ns}\min_{s\geq 0}\sum_{k=1}^{m}\sum_{b\in\mathbb{B}}P_{B_{k}}(b)\sum_{n=1}^{\Ns}\log_{2}\Bigl(1+e^{s(-1)^{b}\lambda^{(n)}_{k,b}}\Bigr)
\end{align}
where $\lambda^{(n)}_{k,b}$, $n=1,2,\ld,\Ns$ are i.i.d. random variables distributed according to the PDF of the L-values $f_{L_{k}|B_{k}}(\lambda|b)$ and $H_{\tnr{b}}(p)\triangleq -p\log_{2}(p)-(1-p)\log_{2}(1-p)$ is the binary entropy function. The maximization over $s$ in \eqref{GMI.general.MC} can be easily approximated (numerically) using the concavity of the GMI on $s$\cite[eq.~(4.81)]{Alvarado15_Book}.

We emphasize here that the expression in \eqref{GMI.general.MC} is valid for any symbol-wise metric in the form of \eqref{metric.BW}, i.e., for any L-value $L_{k}$ that ignores the dependency between the bits in the symbol. In particular, when the L-values are calculated exactly using \eqref{LLR.general}, the GMI can be estimated using \eqref{GMI.general.MC} and $s=1$, which follows from \cite[Theorem~4.20]{Alvarado15_Book}.

\subsection{AWGN Channel}\label{Sec:Prel.AWGN}

Often, if not always, CM transceivers in optical communication systems assume that the discrete-time channel, including transmitter- and receiver-side DSP, is a memoryless AWGN channel $Y=X+Z$, where $Z$ is a complex, zero-mean, circularly symmetric Gaussian random variable with total variance $\Ex[|Z|]^2$.  This assumption might be suboptimal, but in the absence of a better (non-Gaussian) model with memory, the memoryless AWGN channel assumption is reasonable. In this subsection, we specialize the MI and GMI estimators in \eqref{MI.general.MC} and \eqref{GMI.general.MC} to the AWGN channel and equally likely input bits (and therefore, equally likely symbols in $\mcX$).

For the AWGN channel and a uniform input distribution, the MI in \eqref{MI.general} can be estimated using \eqref{MI.general.MC} as
\begin{align}
I(X;Y) &\approx \log_{2}(M)
\label{MI.MC} -\frac{1}{M\Ns}\sum_{i=1}^M\sum_{l=1}^{\Ns} \log_{2}  f_{i,l},
\end{align}
where
\begin{align}
f_{i,l} \triangleq \sum_{j=1}^M \exp{\bigl(-\rho(2\Re\set{(x_{i}-x_{j})^{*} z^{(l)}}+|z^{(l)}|^{2})\bigr)},
\end{align}
the signal-to-noise ratio (SNR) $\rho$ is defined as $\rho\triangleq \Ex_{X}[|X|^{2}]/\Ex[|Z|]^2$, and $z^{(l)}$ with $l=1,2,\ld,\Ns$ are $\Ns$ independent realizations of the Gaussian random variable $Z$.

L-values may be calculated either exactly or using the max-log approximation. In the first case, the exact L-values in \eqref{LLR.general} are calculated as
\begin{align}\label{LLR.sum-exp.AWGN}
L_{k}	& = \log\frac{\sum_{x\in\mcXko}\exp(-\rho |y-x|^{2})}{\sum_{x\in\mcXkz}\exp(-\rho |y-x|^{2})}
\end{align} 
where we used the uniform input symbol distribution assumption. For given sequences of $m\Ns$ transmitted bits $c_{k,l}$ and $m\Ns$ L-values $\lambda_{k,l}$ computed via \eqref{LLR.sum-exp.AWGN}, for $k=1,\ld,m$ and $l=1,\ld,\Ns$, the GMI in \eqref{GMI.general.MC} can be estimated as
\begin{align}
\label{GMI.sum-exp.MC}
\tnr{GMI}	& \approx m-\frac{1}{\Ns}\sum_{k=1}^{m}\sum_{l=1}^{\Ns}\log_{2}\Bigl(1+e^{(-1)^{c_{k,l}}\lambda_{k,l}}\Bigr).
\end{align}
In the second case, the max-log L-values in \eqref{LLR.max-log} are calculated as
\begin{align}
\label{LLR.max-log.AWGN}
L_{k}	& \approx \rho \left(\min_{x\in\mcXkz}|y-x|^{2}-\min_{x\in\mcXko}|y-x|^{2}\right).
\end{align}
For given sequences of transmitted bits $c_{k,l}$ and max-log L-values $\lambda_{k,l}$ computed via \eqref{LLR.max-log.AWGN}, the GMI can be estimated using \eqref{GMI.general.MC} as
\begin{align}
\label{GMI.max-log.MC}
\tnr{GMI}	& \approx m-\frac{1}{\Ns}\min_{s\geq 0}\sum_{k=1}^{m}\sum_{l=1}^{\Ns}\log_{2}\Bigl(1+e^{(-1)^{c_{k,l}}\lambda_{k,l}}\Bigr).
\end{align}

It is important to note at this point that to calculate the GMI, \eqref{GMI.sum-exp.MC} and \eqref{GMI.max-log.MC} should be used for exact and max-log L-values, respectively. Using \eqref{GMI.sum-exp.MC} for max-log L-values results in a rate lower than the true one, i.e., the minimization over $s$ in \eqref{GMI.max-log.MC} is a mandatory step for approximated L-values.

\section{Post-FEC BER Prediction}\label{Prediction}

In this section, we study the robustness of three different metrics to predict the \mbox{post-FEC} BER of \mbox{SD-FEC}: the \mbox{pre-FEC} BER, the MI, and the GMI. The aim is to find a robust and easy-to-measure metric that can be used to predict the \mbox{post-FEC} BER of a given encoder and decoder pair across different channels. Results for the AWGN channel are shown first, followed by results for the nonlinear optical channel.

\subsection{AWGN Channel}\label{Prediction.AWGN}

\begin{figure}[tbp]
\newcommand{\scale}{0.85}
\newcommand{\scalesmall}{0.55}
\begin{minipage}[c]{\columnwidth}
\centering
\psfrag{SCC}[cc][cc][\scalesmall]{{\fcolorbox{white}{white}{{$6.25\%$ OH \cite{Zhang14}}}}}
\psfrag{R13}[cc][cc][\scalesmall][-25]{{\fcolorbox{white}{white}{{$\Rc=1/3$}}}}
\psfrag{R25}[cc][cc][\scalesmall][-25]{{\fcolorbox{white}{white}{{$\Rc=2/5$}}}}
\psfrag{R12}[cc][cc][\scalesmall][-25]{{\fcolorbox{white}{white}{{$\Rc=1/2$}}}}
\psfrag{R35}[cc][cc][\scalesmall][-25]{{\fcolorbox{white}{white}{{$\Rc=3/5$}}}}
\psfrag{R23}[cc][cc][\scalesmall][-25]{{\fcolorbox{white}{white}{{$\Rc=2/3$}}}}
\psfrag{R34}[cc][cc][\scalesmall][-25]{{\fcolorbox{white}{white}{{$\Rc=3/4$}}}}
\psfrag{R56}[cc][cc][\scalesmall][-25]{{\fcolorbox{white}{white}{{$\Rc=5/6$}}}}
\psfrag{ylabel}[cc][cb][\scale]{$\BERout$}%
\psfrag{QPSK}[cl][cl][\scalesmall]{$4$QAM}%
\psfrag{8QAM}[cl][cl][\scalesmall]{$8$QAM}%
\psfrag{64QAM}[cl][cl][\scalesmall]{$64$QAM}%
\psfrag{256QAMM}[cl][cl][\scalesmall]{$256$QAM}%
\psfrag{xlabel}[cc][cc][\scale]{$\BERin$}%
\includegraphics{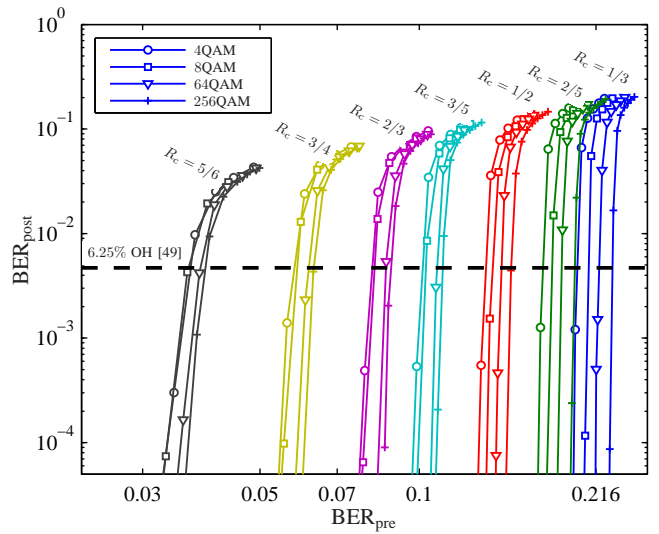}\\
{\small (a) Post-FEC BER vs. \mbox{pre-FEC} BER}\\
\psfrag{xlabel}[cc][cc][\scale]{$I(X;Y)/m$}%
\includegraphics{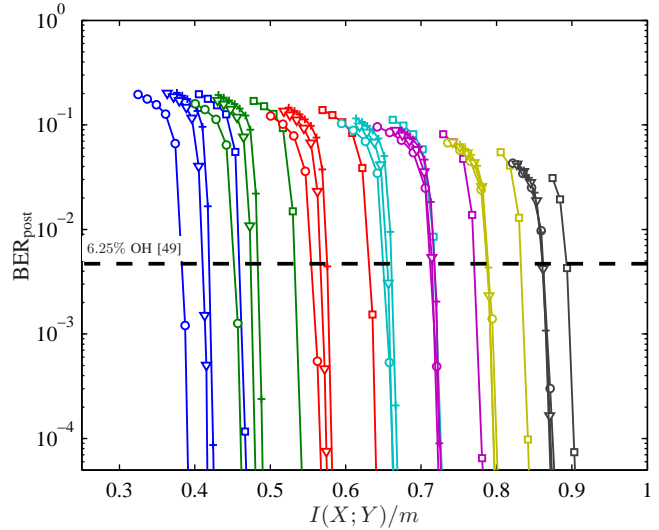}\\
{\small (b) Post-FEC BER vs. normalized MI}\\
\psfrag{xlabel}[cc][cc][\scale]{$\tnr{GMI}/m$}%
\includegraphics{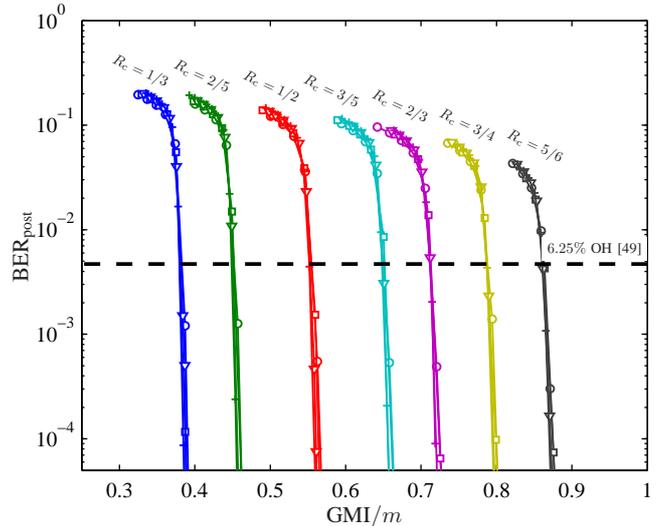}\\
{\small (c) Post-FEC BER vs. normalized GMI}\\
\end{minipage}
\caption{Post-FEC BER for TCs with $\Rc\in\set{2/5,1/2,3/5,2/3,3/4,5/6}$ (colors) and different modulation formats (markers): $4$QAM, $8$QAM, $64$QAM, and $256$QAM. The \mbox{post-FEC} BER is shown versus (a) \mbox{pre-FEC} BER, (b) normalized MI, and (c) normalized GMI. The L-values for $8$QAM and for $256$QAM are calculated using the max-log approximation.}
\label{BERout_TC}
\end{figure}

To study the \mbox{post-FEC} BER prediction \emph{across} different BISO channels (see \figref{model_BISO}~(a)), we consider the TCs defined in \secref{Sec:Prel.Model} and four modulation formats: $M$-QAM constellations with $M=4,8,64,256$. For $M=4,64$, the SD decoder uses exact L-values and for $M=8,256$, max-log L-values\footnote{For $M=256$, the use of max-log L-values is very relevant in practice as the calculation in \eqref{LLR.sum-exp.AWGN} is greatly simplified.}. In \figref{BERout_TC}~(a), the post-LDPC BER is shown as a function of $\BERin$ for the $24$ cases. Ideally, all the lines for the same rate (same color) should fall on top of one another, indicating that measuring $\BERin$ is enough to predict $\BERout$ when the BISO channel (in this case, the modulation format) changes. The results in this figure show that this is not the case, especially for low and medium code rates. The pre-FEC BER therefore fails to predict the performance of the \mbox{SD-FEC} decoder across different BISO channels. 

To estimate the inaccuracy of the \mbox{SD-FEC} limit paradigm, consider the results for $4$QAM and $\Rc=1/3$ shown in \figref{BERout_TC}~(a). For a target post-FEC BER of $\BERout=4.7\cdot 10^{-3}$, the required pre-FEC BER is $\BERin\approx 0.2$. By using the \mbox{SD-FEC} limit paradigm, we can conclude that to guarantee the same for post-FEC BER for $256$QAM, the same pre-FEC BER can be assumed ($\BERin\approx 0.2$). This is clearly not the case, as for $256$QAM and $\Rc=1/3$, the pre-FEC BER can be higher ($\BERin\approx 0.23$). An alternative interpretation of this is that the results in \figref{BERout_TC}~(a) show that for $\BERin\approx 0.2$ and $256$QAM, the code rate can be increased to $\Rc=2/5$. This shows that the use of the the \mbox{SD-FEC} limit paradigm in this scenario leads to an underestimation of the spectral efficiency of $20\%$. Very similar conclusions can be in fact drawn for the LDPC codes shown in \cite[Fig.~4]{Alvarado2015_JLT}. We also conjecture that the use of the \mbox{SD-FEC} limit paradigm in the record results reported in \cite{Beppu15,Qian13ofc} (where a pre-FEC BER threshold obtained for $4$QAM was used for $2048$QAM) are in fact incorrect and even higher spectral efficiencies can be obtained. In this case, however, we expect the underestimation to be below $5\%$.

The results in \figref{BERout_TC}~(a) show the variations on the required pre-FEC BER to guarantee a given post-FEC BER across different modulation formats. While for low code rates these variations could lead to errors of up to $20\%$ in spectral efficiencies, the errors decrease as the code rate increases. This partially suggest that the pre-FEC BER is a relatively good metric for high code rates, however, we have no theoretical justification for the use of  $\BERin$ to predict the performance of a \mbox{SD-FEC}. Furthermore, we believe that having a metric that works for all code rates is important. Considering only high code rates---as is usually done in the optical community---is an artificial constraint that reduces flexibility in the design, as pointed out in \cite[Sec.~II-B]{Smith10}.

\begin{figure}[tbp]
\newcommand{\scale}{0.85}
\newcommand{\scalesmall}{0.6}
\centering
\psfrag{ylabel}[cc][cb][\scale]{PDF}%
\psfrag{xlabel}[cc][cB][\scale]{$l$}%
\psfrag{QAM}[cl][cl][\scalesmall]{$8$QAM}%
\psfrag{256QAM}[cl][cl][\scalesmall]{$256$QAM}%
\includegraphics{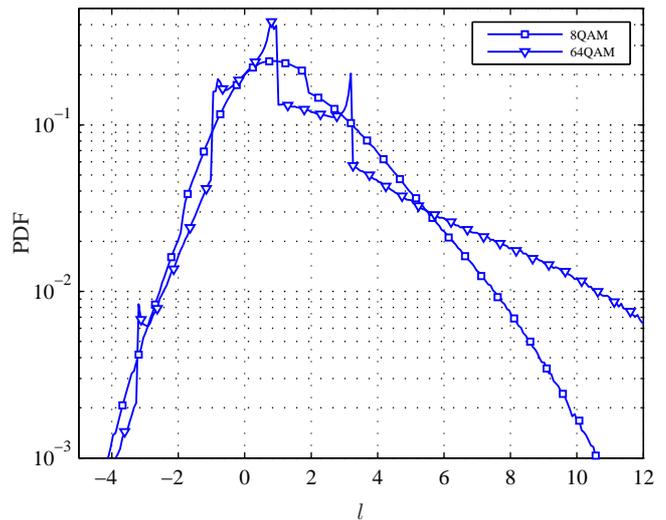}
\caption{Conditional PDF of the L-values $f_{L|B}(l|1)$ in \eqref{GMI.LLRs.Symm.Mix.PDF} for $8$QAM and $64$QAM. In both cases the L-values are calculated using \eqref{LLR.sum-exp.AWGN}.}
\label{LLR_histograms}
\end{figure}

An intuitive explanation for the results in \figref{BERout_TC}~(a) is that the \mbox{SD-FEC} in \figref{model_general_2pol} does not operate on bits, and thus, a metric that is based on bits (i.e., the \mbox{pre-FEC} BER) cannot be used to predict the performance of the decoder. To clarify this, we compare $8$QAM and $64$QAM for $\Rc=1/3$ and a target $\BERin\approx 0.216$. Exact L-value calculations are considered in both cases. From \figref{BERout_TC}~(a) we see that $\BERout\approx 5\cdot 10^{-4}$ for $64$QAM. For $8$QAM, this value is $\BERout\approx 5\cdot 10^{-2}$, which is slightly lower than the one shown in \figref{BERout_TC}~(a) for max-log L-values. In \figref{LLR_histograms} we show the PDF\footnote{Estimated via histograms.} 
\begin{align}\label{GMI.LLRs.Symm.Mix.PDF}
f_{L|B}(l|b) = \frac{1}{2m}\sum_{k=1}^{m} f_{L_{k}|B_{k}}(l|b)+f_{L_{k}|B_{k}}(-l|1-b).
\end{align}
The PDF in \eqref{GMI.LLRs.Symm.Mix.PDF} corresponds to the conditional PDF of ``symmetrized'' and ``mixed'' L-values. For exact L-values, this PDF has been recently shown in \cite[Sec.~V]{Ivanov14_Arxiv} to fully determine the GMI (via $\tnr{GMI} = m I(B;L)$). Under the uniform bit probability assumption, the \mbox{pre-FEC} BER in \eqref{BERin.def.2} can be expressed as
\begin{align}
\label{BERin.mixed.sym.LLR.1}
\BERin 	& =   \frac{1}{2m}\sum_{k=1}^{m} \int_{-\infty}^{0} (f_{L_{k}|B_{k}}(-l | 0)+f_{L_{k}|B_{k}}(l | 1))\,\tnr{d}l
\end{align}
and thus, it is clear that the \mbox{pre-FEC} BER can be calculated by
\begin{align}
\label{BERin.mixed.sym.LLR.2}
\BERin 	& =   \int_{-\infty}^{0} f_{L|B}(l|1)\,\tnr{d}l
\end{align}
where $f_{L|B}(l|b)$ is given by \eqref{GMI.LLRs.Symm.Mix.PDF}.

While both PDFs $f_{L|B}(l|1)$ in \figref{LLR_histograms} give the same \mbox{pre-FEC} BER ($\BERin\approx 0.216$), the \mbox{post-FEC} BER for $64$QAM is much lower than the one for $8$QAM. This can be explained by the different shapes of the PDFs in \figref{LLR_histograms}. In particular, the slow-decaying right tail of the PDF of $64$QAM shows that some L-values with high reliability (i.e., high magnitude) will be observed, which the iterative \mbox{SD-FEC} decoder can exploit.

Using $\BERin$ to predict the performance of \mbox{SD-FEC} decoders has no information-theoretic justification. To remedy this, one could consider the symbol-wise MI $I(X;Y)$ (see \figref{model_general_2pol}) as a metric to better predict $\BERout$. The values of $\BERout$ as a function of the normalized MI $I(X;Y)/m$ are shown in \figref{BERout_TC}~(b). In this case too, the prediction does not work well across all rates. In particular, we note that although for square QAM constellations ($M=4,64,256$) the MI seems to work well for high code rates (as previously reported in \cite[Sec.~III]{Alvarado2015_JLT}), this is not the case if $8$QAM is considered. The MI then appears to be less reliable to predict $\BERout$ than the \mbox{pre-FEC} BER.

One intuitive explanation for the results in \figref{BERout_TC}~(b) is that the MI is an achievable rate for the \emph{optimum} receiver in \eqref{rule.ML}, but not for the (suboptimal) receiver in \figref{model_general_2pol} (see \eqref{rule.BW}). Another explanation is related to the performance dependence of $\BERout$ on the binary labeling of the constellation. It is nowadays well understood that for the receiver in \figref{model_general_2pol}, the performance of the \mbox{SD-FEC} decoder depends on the binary labeling; Gray (or quasi-Gray) labelings are known to be among the best. On the other hand, the MI does not depend on the binary labeling but only on the constellation. Thus, it is not surprising that a labeling-independent metric fails at predicting the labeling-dependent $\BERout$.

The third and last metric we consider to predict $\BERout$ is the GMI. The rationale behind this is that an \mbox{SD-FEC} decoder is fed with L-values, and thus, the GMI (see \eqref{GMI.LLRs}) is an intuitively reasonable metric. The values of $\BERout$ as a function of the normalized GMI are shown in \figref{BERout_TC}~(c). These results show that for a given code rate, changing the constellation does not affect the \mbox{post-FEC} BER prediction based on the GMI. More importantly, and unlike for $\BERout$ and MI, the prediction based on the GMI appears to work across \emph{all} code rates. These results in fact show that the considered TCs appear to be universal (with respect to the GMI), which, to the best of our knowledge, has never been shown in the literature.
    
\begin{figure}[tbp]
\newcommand{\scale}{0.85}
\centering
\psfrag{ylabel2}[cc][cb][\scale]{$I(X;Y)/m$}%
\psfrag{ylabel3}[cc][cb][\scale]{$\tnr{GMI}/m$}%
\psfrag{xlabel}[cc][cc][\scale]{$\Rc$}%
\begin{minipage}[c]{\columnwidth}
\centering
\includegraphics{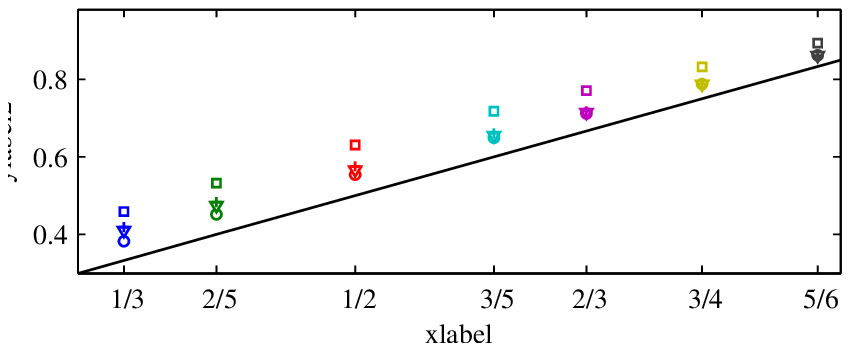}\\
{\small (a) Normalized MI}\\
\includegraphics{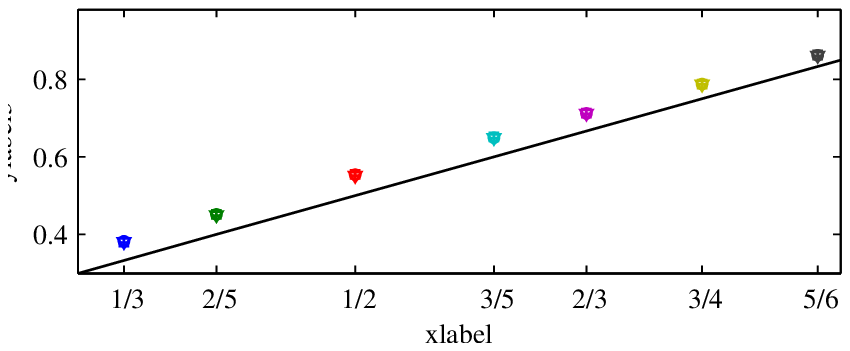}\\
{\small (b) Normalized GMI}\\
\end{minipage}
\caption{Required values for the different metrics to give $\BERout=4.7\cdot 10^{-3}$ as a function of the code rate for the same cases as in \figref{BERout_TC}: (a) normalized MI and (b) normalized GMI. The curves $I(X;Y)=m\Rc$ and $\tnr{GMI}=m\Rc$ are shown in (a) and (b), resp.}
\label{TC_Summary}
\end{figure}

\figref{TC_Summary} shows the values of MI and GMI needed for each configuration in \figref{BERout_TC} to reach a \mbox{post-FEC} BER of $\BERout=4.7\cdot 10^{-3}$.\footnote{Note that similar results could be presented in terms of pre-FEC BER. To have a fair comparison in terms of rates, however, one would need to convert the pre-FEC BER into SNR, and then map that SNR onto MI (or GMI), giving exactly what is shown in \figref{BERout_TC}.} These values are obtained by finding the crossing points of the curves in \figref{BERout_TC} and the horizontal dashed lines. \figref{TC_Summary} also shows the relationships $I(X;Y)=m\Rc$ and $\tnr{GMI}=m\Rc$, where the vertical difference between the markers and the solid lines represent the rate penalty for these codes. The results in \figref{TC_Summary} clearly show the excellent prediction based on GMI and how MI does not work well across different modulation formats. 

We showed before that the pre-FEC BER can lead to an erroneous estimate of the spectral efficiency, which is particularly noticeable for low code rates. A similar problem occurs if the normalized MI in \figref{TC_Summary}~(a) is used to predict post-FEC BER. For example, the results in \figref{TC_Summary}~(a) show that post-FEC BER of $\BERout = 4.7\cdot 10^{-3}$ can be achieved with $4$QAM and $\Rc=2/3$ when the normalized MI is approximately $0.71$ (see also \figref{BERout_TC}~(b)). One might be tempted to then conclude that, for the same MI, the same post-FEC BER can be achieved with $8$QAM and $\Rc=2/3$. The results in \figref{TC_Summary}~(a) show that this is in fact not possible, and a (lower) code rate of $\Rc=3/5$ is needed. In other words, the use of a ``MI threshold paradigm'' could lead to an overestimation (in this case by $11\%$) of the true spectral efficiency. This is not the case for the GMI (see \figref{TC_Summary}~(b)), where all markers for the same code fall on top of one another.

\subsection{Optical Channel---Simulations}\label{Prediction.NLSE}

Dual-polarization transmission over the nonlinear optical channel specified in \secref{Sec:Prel.Model} was simulated using the coupled polarization nonlinear Schr\"{o}dinger equation (NLSE) \cite[eq.~(6)]{Menyuk87}. This enabled the consideration of an idealized transmission link with zero polarization mode dispersion. The simulations were carried out via the split-step Fourier method with a step size of $100~\tnr{m}$ and an oversampling factor of $4$~\tnr{samples/symbol}.

\begin{figure}[tbp]
\newcommand{\scale}{0.85}
\newcommand{\scalesmall}{0.6}
\centering
\psfrag{ylabel}[cc][cb][\scale]{Achievable Rate~[bit/symbol]}%
\psfrag{xlabel}[cc][cB][\scale]{Distance $L$~[km]}%
\psfrag{QPSK}[cl][cl][\scalesmall]{$4$QAM}%
\psfrag{16QAM}[cl][cl][\scalesmall]{$16$QAM}%
\psfrag{64QAM}[cl][cl][\scalesmall]{$64$QAM}%
\psfrag{256QAM}[cl][cl][\scalesmall]{$256$QAM}%
\includegraphics{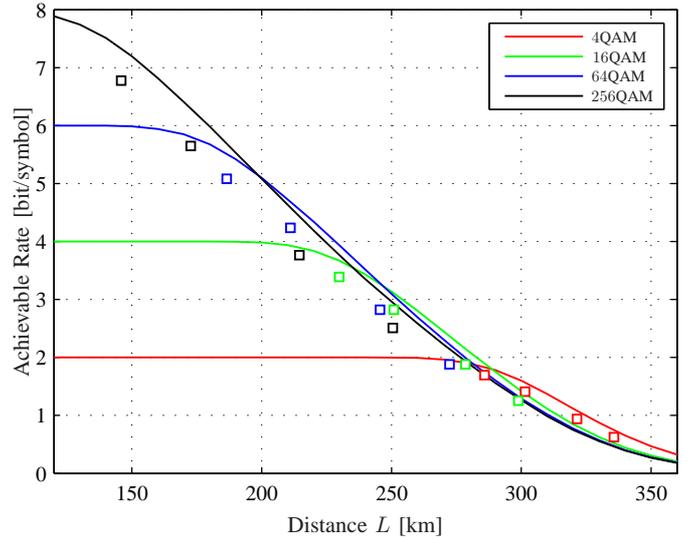}
\caption{Achievable rates (per polarization) versus span length: GMI (solid lines) and LDPC codes with $\Rc\in\set{1/3, 1/2, 3/4, 9/10}$ (markers).}
\label{Rates_vs_L}
\end{figure}

\figref{Rates_vs_L} shows the GMI (per polarization) as a function of the span length, for $M$QAM constellations with $M=4,16,64,256$. For each distance and $M$, we used the launch power that gave the highest GMI. In this figure, we also show the distance required by the LDPC codes in \secref{Sec:Prel.Model} to give $\BERout = 4.7\cdot 10^{-3}$ for each combination of four constellations and $\Rc\in\set{1/3, 1/2, 3/4, 9/10}$. The vertical position of these $16$ markers represent the resulting achievable rates and clearly show that the results follow the GMI curves. This is in good agreement with the results in \cite{Alvarado13c,Alvarado2015_JLT}, where it was shown that the GMI can be used to predict the performance of LDPC codes for the AWGN channel. The penalties with respect to the GMI are between $5$ and $15$ km and are highest for high code rates and large values of $M$. These penalties are  caused by the suboptimality of the LDPC code under consideration.

\begin{figure}[tbp]
\newcommand{\scale}{0.85}
\newcommand{\scalesmall}{0.55}
\begin{minipage}[c]{\columnwidth}
\centering
\psfrag{SCC}[cc][cc][\scalesmall]{{\fcolorbox{white}{white}{{$6.25\%$ OH \cite{Zhang14}}}}}
\psfrag{R13}[cr][cr][\scalesmall][-20]{{\fcolorbox{white}{white}{{$\Rc=1/3$}}}}
\psfrag{R12}[cr][cr][\scalesmall][-20]{{\fcolorbox{white}{white}{{$\Rc=1/2$}}}}
\psfrag{R34}[cr][cr][\scalesmall][-20]{{\fcolorbox{white}{white}{{$\Rc=3/4$}}}}
\psfrag{R91}[cr][cr][\scalesmall][-20]{{\fcolorbox{white}{white}{{$\Rc=9/10$}}}}
\psfrag{ylabel}[cc][cb][\scale]{$\BERout$}%
\psfrag{QPSK}[cl][cl][\scalesmall]{$4$QAM (NLSE)}%
\psfrag{16QAM}[cl][cl][\scalesmall]{$16$QAM (NLSE)}%
\psfrag{64QAM}[cl][cl][\scalesmall]{$64$QAM (NLSE)}%
\psfrag{256QAMM}[cl][cl][\scalesmall]{$256$QAM (NLSE)}%
\psfrag{QPSKAWGN}[cl][cl][\scalesmall]{$4$QAM (AWGN)}%
\psfrag{16QAMAWGN}[cl][cl][\scalesmall]{$16$QAM (AWGN)}%
\psfrag{64QAMAWGN}[cl][cl][\scalesmall]{$64$QAM (AWGN)}%
\psfrag{256QAMMAWGN}[cl][cl][\scalesmall]{$256$QAM (AWGN)}%
\psfrag{xlabel}[cc][cc][\scale]{$\BERin$}%
\includegraphics{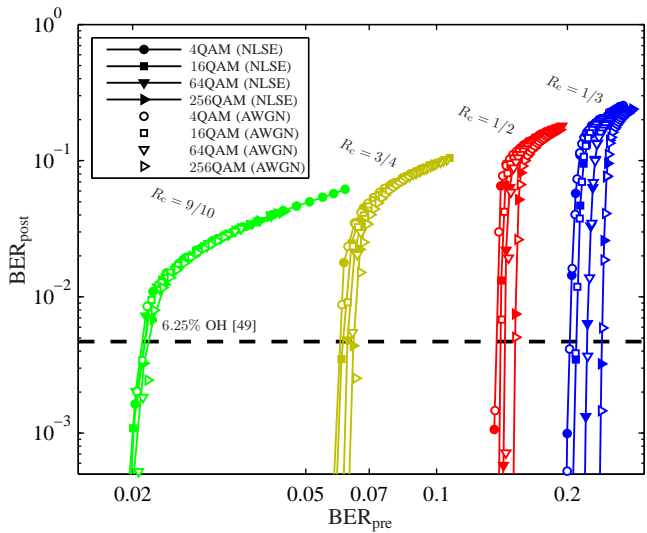}\\
{\small (a) Post-FEC BER vs. \mbox{pre-FEC} BER}\\
\psfrag{xlabel}[cc][cc][\scale]{$I(X;Y)/m$}%
\includegraphics{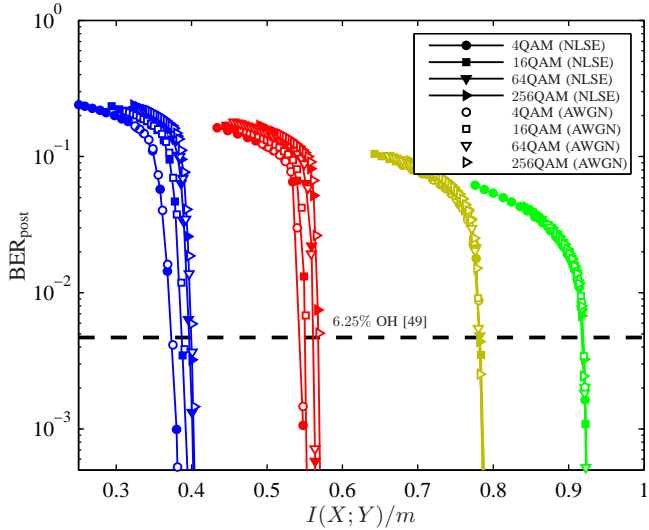}\\
{\small (b) Post-FEC BER vs. normalized MI}\\
\psfrag{xlabel}[cc][cc][\scale]{$\tnr{GMI}/m$}%
\includegraphics{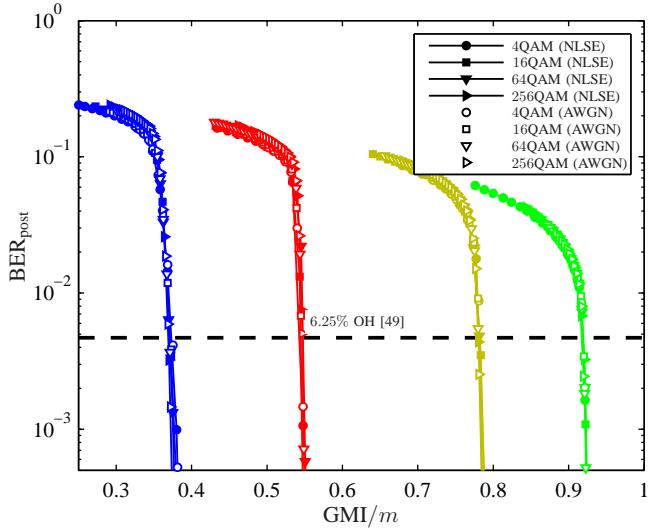}\\
{\small (c) Post-FEC BER vs. normalized GMI}\\
\end{minipage}
\caption{Post-FEC BER for LPDC codes with $\Rc\in\set{1/3, 1/2, 3/4, 9/10}$ (colors) over linear and nonlinear channels and different modulation formats (markers): $4$QAM, $16$QAM, $64$QAM, and $256$QAM. The \mbox{post-FEC} BER is shown versus (a) \mbox{pre-FEC} BER, (b) normalized MI, and (c) normalized GMI. All the L-values are calculated using \eqref{LLR.sum-exp.AWGN}.}
\label{BERout_LDPC}
\end{figure}

In analogy with \figref{BERout_TC}, \figref{BERout_LDPC} shows the \mbox{post-FEC} BER as a function of (a) \mbox{pre-FEC} BER, (b) normalized MI, and (c) normalized GMI. The results for the NLSE are shown with filled markers and show that the prediction based on the GMI is excellent. Just as for TCs, the prediction based on \mbox{pre-FEC} BER does not always work, however, a relatively good approximation is obtained for high code rates.

When compared to the results in \cite[Fig.~5]{Alvarado15a}, we note that the curves in \figref{BERout_LDPC}~(c) are  more ``compact'' for low rates. The difference between the simulation setup in \cite{Alvarado15a} and the one in this paper is that here we consider a random interleaver between the binary encoder and the mapper. Using this interleaver is thus important to make the GMI-based prediction even more precise.

In \figref{BERout_LDPC}, we also show results obtained for the AWGN channel (white markers). These results were obtained for the same modulation and coding pairs as used in the NLSE simulations and show that indeed the GMI is a robust metric to predict \mbox{post-FEC} BER across different channels. In particular, \figref{BERout_LDPC}~(c) shows that the \mbox{post-FEC} BER predictions give the same results for both the AWGN channel and the simulations based on the NLSE. This also suggests that using a Gaussian model for the noise is quite reasonable.

\begin{figure}[tbp]
\newcommand{\scale}{0.85}
\newcommand{\scalesmall}{0.55}
\centering
\psfrag{ylabel}[cc][cb][\scale]{BER}%
\psfrag{xlabel}[cc][cB][\scale]{Launch Power~[dBm]}%
\psfrag{BERin}[cl][cl][\scalesmall]{$\BERin$}%
\psfrag{BERoutLinearRegi}[cl][cl][\scalesmall]{$\BERout$ (linear)}%
\psfrag{BERoutNLinearRegi}[cl][cl][\scalesmall]{$\BERout$ (nonlinear)}%
\psfrag{SCC}[cc][cc][\scalesmall]{{\fcolorbox{white}{white}{{$6.25\%$ OH \cite{Zhang14}}}}}
\includegraphics{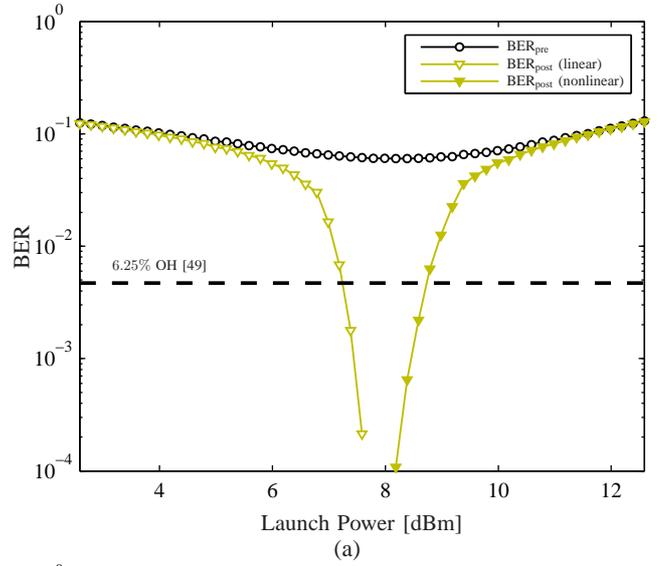}\\
{\small (a) }
\psfrag{ylabel}[cc][cb][\scale]{$\BERout$}%
\psfrag{xlabel}[cc][cB][\scale]{$\tnr{GMI}/m$}%
\psfrag{LinearRegi}[cl][cl][\scalesmall]{Linear}%
\psfrag{NLinearRegi}[cl][cl][\scalesmall]{Nonlinear}%
\psfrag{SCC}[cc][cc][\scalesmall]{{\fcolorbox{white}{white}{{$6.25\%$ OH \cite{Zhang14}}}}}
\includegraphics{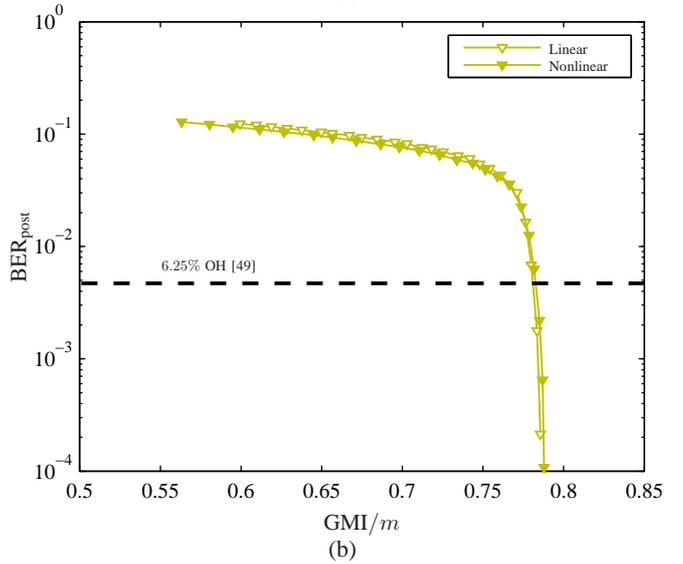}\\
{\small (b) }
\caption{(a) Pre-FEC BER and \mbox{post-FEC} BER as a function of the launch power for an LDPC code with $\Rc=3/4$, $64$QAM, and $L=210$~km. (b) Pre-FEC BER vs. normalized GMI in the linear (white markers) and nonlinear (filled markers) regimes.}
\label{Nonlinear_BER}
\end{figure}

All the results in \figref{BERout_LDPC}~(c) for the NLSE were obtained for the optimal launch power. To show that the GMI prediction is also not dependent on the launch power, we study a fixed distance and vary the launch power, bringing the system deep into the nonlinear regime. As the modulation format, we choose $64$QAM and based on the results in \figref{Rates_vs_L}, we use $L=210$~km and $\Rc=3/4$. The launch power was varied from $2.6$~dBm to $12.6$~dBm, giving the \mbox{pre-FEC} and \mbox{post-FEC} BER shown in \figref{Nonlinear_BER}~(a). The same \mbox{post-FEC} BER values are shown in \figref{Nonlinear_BER}~(b) as a function of the normalized GMI. This figure shows once again that the GMI can be used to accurately predict the \mbox{post-FEC} BER of \mbox{SD-FEC} decoders, even when the channel is highly nonlinear.

\subsection{Optical Channel---Experiments}\label{Prediction.Experiments}

\begin{figure}[tbp]
\centering
\includegraphics[width=\columnwidth]{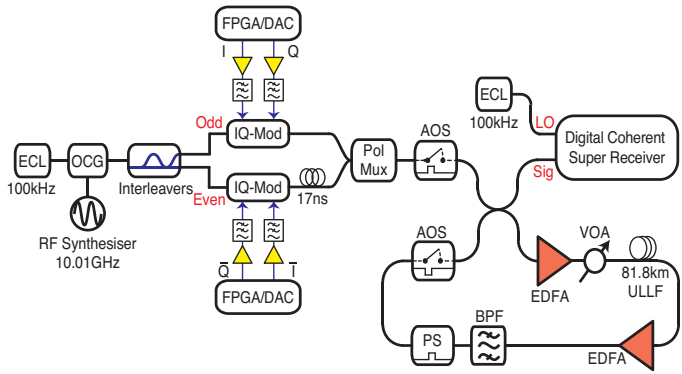}
\caption{$64$QAM Nyquist spaced WDM transmission testbed.}
\label{ExpSetup}
\end{figure}

To experimentally verify that the normalized GMI is an accurate predictor for post-FEC BER, the LDPC code described in Sec.~\ref{Sec:Prel.Model:HD-FEC} was implemented in a dual-polarization $64$QAM Nyquist-spaced WDM transmission system. The corresponding experimental setup is illustrated in \figref{ExpSetup}. A $100~\tnr{kHz}$ linewidth external cavity laser (ECL) was passed through an optical comb generator (OCG) to obtain seven frequency-locked comb lines with a channel spacing of $10.01~\tnr{GHz}$. The eight-level drive signals required for $64$QAM were generated offline in Matlab and were digitally filtered using an RRC filter with a roll-off factor of $0.1$\%. The resulting in-phase (I) and quadrature (Q) signals were loaded onto a pair of field-programmable gate arrays (FPGAs) and output using two digital-to-analog converters (DACs) operating at $20~\tnr{Gsamples/s}$ ($2$~\tnr{samples/symbol}). The odd and even sub-carriers were independently modulated using two complex IQ modulators, which were subsequently decorrelated before being combined and polarization multiplexed to form a Nyquist spaced  $64$QAM super-carrier. The recirculating loop configuration consisted of two acousto-optic switches (AOS), two EDFAs with a noise figure of $4.5~\tnr{dB}$, an optical band-pass filter (BPF) for amplified spontaneous emission noise removal, a loop-synchronous polarization scrambler (PS) and a single $81.8~\tnr{km}$ span of Corning$^{\circledR}$ SMF-28$^{\circledR}$ ULL fiber.

The polarization-diverse coherent receiver had an electrical bandwidth of $70~\tnr{GHz}$ and used a second $100~\tnr{kHz}$ linewidth ECL as a local oscillator (LO). The frequency of the LO was set to coincide with the central sub-carrier of the $64$QAM super-carrier and the received signals were captured using a $160~\tnr{Gsamples/s}$ real-time sampling oscilloscope with $63~\tnr{GHz}$ analog electrical bandwidth. DSP and \mbox{SD-FEC} decoding were subsequently performed offline in Matlab and was identical to that described in \cite{Maher15}.

\begin{figure}[tbp]
\centering
\newcommand{\scale}{0.85}
\newcommand{\scalesmall}{0.55}
\newcommand{\scalesmallsmall}{0.5}
\psfrag{ylabel}[cc][cb][\scale]{$\BERout$}%
\psfrag{xlabel}[cc][cB][\scale]{$\tnr{GMI}/m$}%
\psfrag{000}[cl][cl][\scalesmall]{B2B}%
\psfrag{80}[cl][cl][\scalesmall]{$\Nspan=1$}%
\psfrag{320}[cl][cl][\scalesmall]{$\Nspan=4$}%
\psfrag{640}[cl][cl][\scalesmall]{$\Nspan=8$}%
\psfrag{1280---}[cl][cl][\scalesmall]{$\Nspan=16$}%
\psfrag{R25}[cl][cl][\scalesmall][-88]{{\fcolorbox{white}{white}{{$\Rc=2/5$}}}}
\psfrag{R12}[cl][cl][\scalesmall][-88]{{\fcolorbox{white}{white}{{$\Rc=1/2$}}}}
\psfrag{R35}[cl][cl][\scalesmall][-88]{{\fcolorbox{white}{white}{{$\Rc=3/5$}}}}
\psfrag{R34}[cl][cl][\scalesmall][-88]{{\fcolorbox{white}{white}{{$\Rc=3/4$}}}}
\psfrag{R910}[cl][cl][\scalesmall][-88]{{\fcolorbox{white}{white}{{$\Rc=9/10$}}}}
\psfrag{SCC}[cc][cc][\scalesmall]{{\fcolorbox{white}{white}{{$6.25\%$ OH \cite{Zhang14}}}}}
\psfrag{B2B}[cl][cl][\scalesmallsmall][30]{}
\psfrag{R25N16P1}[cr][cr][\scalesmallsmall][30]{}
\psfrag{R25N16P2}[cr][cr][\scalesmallsmall][30]{}
\psfrag{R12N16P1}[cl][cl][\scalesmallsmall][30]{}
\psfrag{R12N16P2}[cl][cl][\scalesmallsmall][30]{}
\psfrag{R12N16P3}[cr][cr][\scalesmallsmall][30]{}
\psfrag{R12N16P4}[cl][cl][\scalesmallsmall][30]{}
\psfrag{R35N8P1}[cl][cl][\scalesmallsmall][30]{}
\psfrag{R35N8P2}[cl][cl][\scalesmallsmall][30]{}
\psfrag{R35N8P3}[cl][cl][\scalesmallsmall][30]{}
\psfrag{R35N4P4}[cl][cl][\scalesmallsmall][30]{}
\psfrag{R34N4P1}[cl][cl][\scalesmallsmall][30]{}
\psfrag{R34N4P2}[cl][cl][\scalesmallsmall][30]{}
\psfrag{R34N4P3}[cr][cr][\scalesmallsmall][30]{}
\psfrag{R34N4P4}[cr][cr][\scalesmallsmall][30]{}
\psfrag{R34N4P5}[cl][cl][\scalesmallsmall][30]{}
\psfrag{R910N4P1}[cl][cl][\scalesmallsmall][45]{}
\psfrag{R910N4P2}[cl][cl][\scalesmallsmall][45]{}
\psfrag{R910N4P3}[cl][cl][\scalesmallsmall][45]{}
\psfrag{R910N4P4}[cl][cl][\scalesmallsmall][45]{}
\psfrag{R910N4P5}[cl][cl][\scalesmallsmall][45]{}
\includegraphics{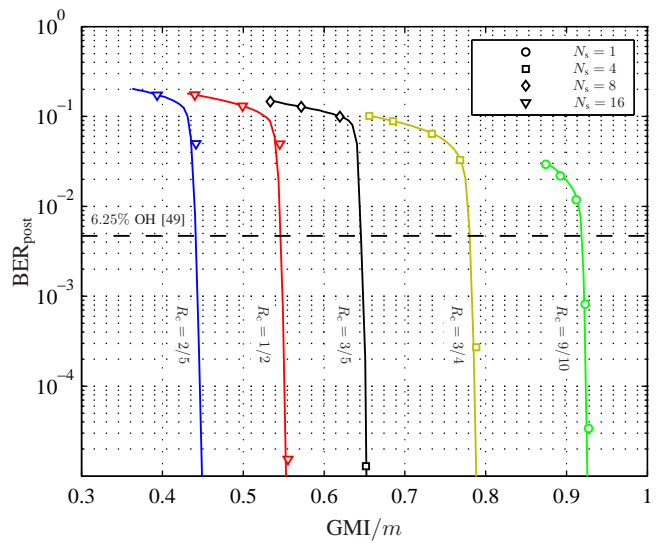}
\caption{Post-FEC BER for  the $64$QAM Nyquist spaced WDM transmission testbed with LPDC codes and $\Rc\in\set{2/5,1/2,3/5,3/4,9/10}$ (colors) as a function of the normalized GMI. Experimental results for different number of spans $\Nspan$ are shown with markers and AWGN results with solid lines.}
\label{BERout_vs_GMI_LDPC_Lab_w_interleaver}
\end{figure}

The transmission performance of the central WDM carrier was analyzed over a number of transmission distances from $81.8~\tnr{km}$ ($\Nspan=1$) to $1308.8~\tnr{km}$ ($\Nspan=16$) and for a number of launch powers, ranging from $-18~\tnr{dBm}$ to $+2~\tnr{dBm}$. This resulted in a normalized GMI ranging from $0.39$ to $0.93$, which required adaptation of the OH in order to achieve a post-FEC BER that was below the target BER after \mbox{SD-FEC} decoding $\BERout=4.7\cdot 10^{-3}$. \figref{BERout_vs_GMI_LDPC_Lab_w_interleaver} illustrates the experimentally measured normalized GMI (markers) as a function of post-FEC BER, for five code rates $\Rc\in\set{2/5,1/2,3/5,3/4,9/10}$. The transmission distances were $81.8, 327.2, 654.4$ and $1308.8~\tnr{km}$, i.e., $\Nspan=1,4,8,$ and $16$ spans. The simulated results obtained for an AWGN channel (i.e., the ones in \figref{BERout_LDPC}~(c)) are displayed using solid lines. Excellent agreement between the simulated curves and the experimental points is demonstrated for all \mbox{SD-FEC} code rates, launch powers, and distances, even though the simulations and experiments concern entirely different channels.\footnote{Note also that the parameters of the experimental setup in this section are different to those in \secref{Prediction.NLSE}.}

Each result shown with a marker in \figref{BERout_vs_GMI_LDPC_Lab_w_interleaver} corresponds to a given launch power (per channel), code rate $\Rc$, and number of spans $\Nspan$. These results are summarized in Table~\ref{tab:lab_res}, where the launch powers are also shown. The results in \figref{BERout_vs_GMI_LDPC_Lab_w_interleaver} and Table~\ref{tab:lab_res} show that regardless of the transmit power, the normalized GMI can indeed be used to predict the post-FEC BER. These results can be seen as experimental validation of those presented in \figref{Nonlinear_BER}.
 
\begin{table}[thp]
	\caption{Summary of results for the experimental setup in \figref{ExpSetup}. Each row corresponds to a marker in \figref{BERout_vs_GMI_LDPC_Lab_w_interleaver}.}
	\centering
	\begin{tabular}{ l  c  c  c c}
\hline

\hline
	Launch Power 				& $\tnr{GMI}/m$ &	$\BERout$	& Spans $\Nspan$	& Rate $\Rc$\\
\hline

\hline 
	$-18.27~\tnr{dBm}$				& $0.39$	& $1.7\cdot 10^{-1}$	& $16$& \multirow{2}{*}{${2}/{5}$}	\\
	$-17.00~\tnr{dBm}$				& $0.44$	& $5.0\cdot 10^{-2}$	& $16$\\
\hline
	$-17.00~\tnr{dBm}$				& $0.44$	& $1.7\cdot 10^{-1}$	& $16$& \multirow{4}{*}{${1}/{2}$}	\\
	$-15.90~\tnr{dBm}$				& $0.50$	& $1.3\cdot 10^{-1}$	& $16$\\
	$-14.80~\tnr{dBm}$				& $0.55$	& $4.9\cdot 10^{-2}$	& $16$\\
	$-13.69~\tnr{dBm}$				& $0.56$	& $1.5\cdot 10^{-5}$	& $16$\\
\hline
	$-18.12~\tnr{dBm}$				& $0.53$	& $1.5\cdot 10^{-2}$	& $8$& \multirow{4}{*}{${3}/{5}$}	\\
	$-17.14~\tnr{dBm}$				& $0.57$	& $1.3\cdot 10^{-2}$	& $8$\\
	$-15.94~\tnr{dBm}$				& $0.61$	& $9.9\cdot 10^{-3}$	& $8$\\
	$-18.21~\tnr{dBm}$				& $0.64$	& $1.3\cdot 10^{-5}$	& $4$\\
\hline
	$-18.20~\tnr{dBm}$				& $0.66$	& $1.0\cdot 10^{-1}$	& $4$& \multirow{5}{*}{${3}/{4}$}	\\
	$-17.20~\tnr{dBm}$				& $0.69$	& $8.8\cdot 10^{-2}$	& $4$\\
	$-16.01~\tnr{dBm}$				& $0.73$	& $6.4\cdot 10^{-2}$	& $4$\\
	$-15.0~\tnr{dBm}$				& $0.77$	& $3.3\cdot 10^{-2}$	& $4$\\
	$-0.60~\tnr{dBm}$				& $0.78$	& $2.7\cdot 10^{-4}$	& $4$\\
\hline
	$-9.17~\tnr{dBm}$				& $0.87$	& $2.9\cdot 10^{-2}$	& $1$& \multirow{5}{*}{${9}/{10}$}	\\
	$-10.17~\tnr{dBm}$				& $0.89$	& $2.2\cdot 10^{-2}$	& $1$\\
	$-11.24~\tnr{dBm}$				& $0.91$	& $1.2\cdot 10^{-2}$	& $1$\\
	$-12.24~\tnr{dBm}$				& $0.92$	& $8.2\cdot 10^{-4}$	& $1$\\
	$-13.30~\tnr{dBm}$				& $0.93$	& $3.4\cdot 10^{-5}$	& $1$\\
\hline

\hline
	\end{tabular}
	\label{tab:lab_res}
\end{table}

\section{Conclusions}\label{Conclusions}

This paper studied the GMI as a powerful tool to predict the \mbox{post-FEC} BER of soft-decision FEC. The GMI was measured in experiments and simulations, and for all the considered scenarios proved to be very robust. The GMI can be used to predict the \mbox{post-FEC} BER without actually encoding and decoding data. 

The \mbox{pre-FEC} BER and MI were also shown to be weak predictors of the performance of soft-decision FEC for bit-wise decoders. The so-called FEC limit is, hence, an unreliable design criterion for optical communication systems with soft-decision FEC. On the other hand, the GMI was found to give very good results for all code rates, all considered modulation formats, LDPC and turbo codes, exact and approximated L-values, and for both linear and nonlinear optical transmission. We suggest to replace the ``SD-FEC limit'' (used for many years with hard-decision decoding and now becoming increasingly popular with soft decision) with a ``GMI limit'', which is relevant for modern optical communication systems.

This paper considered only noniterative binary decoding. Different results are expected if a (capacity-approaching soft-decision) nonbinary decoder or a binary decoder with iterative detection (i.e., with soft information being exchanged iteratively between the decoder and demapper) are used. In these cases, we conjecture the MI to be the correct metric to predict the \mbox{post-FEC} BER. This comparison is left for future work.

\balance    

\section*{Acknowledgments}\label{Sec:Ack}

The authors would like to thank Mikhail Ivanov and Christian H\"ager (Chalmers University of Technology) for fruitful discussions regarding the relationship between the GMI and L-values, and Prof. Leszek Szczecinski (INRS-EMT) and Dr. Laurent Schmalen (Alcatel-Lucent Bell Labs) for fruitful discussions regarding post-FEC BER prediction. The authors would also like to thank Dr. Sergejs Makovejs and Corning Inc. for supplying the fiber used in the experimental setup.

\bibliography{IEEEabrv,references_GMI_paper}

\begin{thebibliography}{10}
\providecommand{\url}[1]{#1}
\csname url@samestyle\endcsname
\providecommand{\newblock}{\relax}
\providecommand{\bibinfo}[2]{#2}
\providecommand{\BIBentrySTDinterwordspacing}{\spaceskip=0pt\relax}
\providecommand{\BIBentryALTinterwordstretchfactor}{4}
\providecommand{\BIBentryALTinterwordspacing}{\spaceskip=\fontdimen2\font plus
\BIBentryALTinterwordstretchfactor\fontdimen3\font minus
  \fontdimen4\font\relax}
\providecommand{\BIBforeignlanguage}[2]{{%
\expandafter\ifx\csname l@#1\endcsname\relax
\typeout{** WARNING: IEEEtran.bst: No hyphenation pattern has been}%
\typeout{** loaded for the language `#1'. Using the pattern for}%
\typeout{** the default language instead.}%
\else
\language=\csname l@#1\endcsname
\fi
#2}}
\providecommand{\BIBdecl}{\relax}
\BIBdecl

\bibitem{Fabregas08_Book}
A.~{Guill\'en i F\`abregas}, A.~Martinez, and G.~Caire, ``Bit-interleaved coded
  modulation,'' \emph{Foundations and Trends in Communications and Information
  Theory}, vol.~5, no. 1--2, pp. 1--153, 2008.

\bibitem{Alvarado15_Book}
L.~Szczecinski and A.~Alvarado, \emph{Bit-Interleaved Coded Modulation:
  Fundamentals, Analysis and Design}.\hskip 1em plus 0.5em minus 0.4em\relax
  John Wiley \& Sons, 2015.

\bibitem{Zehavi92}
E.~Zehavi, ``8-{PSK} trellis codes for a {Rayleigh} channel,'' \emph{IEEE
  Trans. Commun.}, vol.~40, no.~3, pp. 873--884, May 1992.

\bibitem{Caire98}
G.~Caire, G.~Taricco, and E.~Biglieri, ``Bit-interleaved coded modulation,''
  \emph{IEEE Trans. Inf. Theory}, vol.~44, no.~3, pp. 927--946, May 1998.

\bibitem{Djordjevic2006_JSQE}
I.~B. Djordjevic, S.~Sankaranarayanan, S.~K. Chilappagari, and B.~Vasic,
  ``Low-density parity-check codes for 40-{G}b/s optical transmission
  systems,'' \emph{IEEE J. Quantum Electron.}, vol.~12, no.~4, pp. 555--562,
  July/Aug. 2006.

\bibitem{Bulow2009}
H.~B\"{u}low and T.~Rankl, ``Soft coded modulation for sensitivity enhancement
  of coherent {100-Gbit/s} transmission systems,'' in \emph{Proc. Optical Fiber
  Communication Conference (OFC)}, San Diego, CA, Mar. 2009.

\bibitem{Bulow2011b}
H.~B\"{u}low, {\"{U}. Abay}, A.~Schenk, and J.~B. Huber, ``Coded modulation of
  polarization- and space-multiplexed signals,'' in \emph{Asia Communications
  and Photonics Conference and Exhibition (ACP)}, Shanghai, China, Nov. 2011.

\bibitem{Millar14}
D.~S. Millar, T.~Koike-Akino, R.~Maher, D.~Lavery, M.~Paskov, K.~Kojima,
  K.~Parsons, B.~C. Thomsen, S.~J. Savory, and P.~Bayvel, ``Experimental
  demonstration of 24-dimensional extended {Golay} coded modulation with
  {LDPC},'' in \emph{Proc. Optical Fiber Communication Conference (OFC)}, San
  Francisco, CA, Mar. 2014.

\bibitem{Hager14a}
C.~H\"{a}ger, A.~{Graell i Amat}, F.~Br\"{a}nnstr\"{o}m, A.~Alvarado, and
  E.~Agrell, ``Improving soft {FEC} performance for higher-order modulations
  via optimized bit channel mappings,'' \emph{Opt. Express}, vol.~22, no.~12,
  pp. 14\,544--14\,558, June 2014.

\bibitem{Alvarado13c}
A.~Alvarado and E.~Agrell, ``Achievable rates for four-dimensional coded
  modulation with a bit-wise receiver,'' in \emph{Proc. Optical Fiber
  Communication Conference (OFC)}, San Francisco, CA, Mar. 2014.

\bibitem{Li97}
X.~Li and J.~A. Ritcey, ``Bit-interleaved coded modulation with iterative
  decoding,'' \emph{IEEE Commun. Lett.}, vol.~1, no.~6, pp. 169--171, Nov.
  1997.

\bibitem{Brink98}
S.~ten Brink, J.~Speidel, and R.-H. Yan, ``Iterative demapping for {QPSK}
  modulation,'' \emph{IEE Electronics Letters}, vol.~34, no.~15, pp.
  1459--1460, July 1998.

\bibitem{Benedetto98b}
S.~Benedetto, G.~Montorsi, D.~Divsalar, and F.~Pollara, ``Soft-input
  soft-output modules for the construction and distributed iterative decoding
  of code networks,'' \emph{Eur. Trans. on Telecommun.}, vol.~9, no.~2, pp.
  155--172, Mar.--Apr. 1998.

\bibitem{Djordjevic2007_JLT}
I.~B. Djordjevic, M.~Cvijetic, L.~Xu, and T.~Wang, ``Using {LDPC}-coded
  modulation and coherent detection for ultra highspeed optical transmission,''
  \emph{J. Lightw. Technol.}, vol.~25, no.~11, pp. 3619--3625, Nov. 2007.

\bibitem{Batshon2009_JLT}
H.~B. Batshon, I.~B. Djordjevic, L.~Xu, and T.~Wang, ``Multidimensional
  {LDPC-Coded} modulation for beyond {400 Gb/s} per wavelength transmission,''
  \emph{IEEE Photon. Technol. Lett.}, vol.~21, no.~16, pp. 1139--1141, Aug.
  2009.

\bibitem{Bulow14}
H.~Buelow, X.~Lu, L.~Schmalen, A.~Klekamp, and F.~Buchali, ``Experimental
  performance of {4D} optimized constellation alternatives for {PM-8QAM} and
  {PM-16QAM},'' in \emph{Proc. Optical Fiber Communication Conference (OFC)},
  San Francisco, CA, Mar. 2014.

\bibitem{Bulow2011}
H.~B\"{u}low and E.~Masalkina, ``Coded modulation in optical communications,''
  in \emph{Proc. Optical Fiber Communication Conference (OFC)}, Los Angeles,
  CA, Mar. 2011.

\bibitem{Schmalen14}
L.~Schmalen, ``Energy efficient {FEC} for optical transmission systems,'' in
  \emph{Proc. Optical Fiber Communication Conference (OFC)}, San Francisco, CA,
  Mar. 2014.

\bibitem{Puc99}
A.~Puc, F.~Kerfoot, A.~Simons, and D.~L. Wilson, ``Concatenated {FEC}
  experiment over 5000 km long straight line {WDM} test bed,'' in \emph{Proc.
  Optical Fiber Communication Conference (OFC)}, San Diego, CA, Feb. 1999.

\bibitem{Ait-Sab00}
O.~{Ait Sab} and V.~Lemaire, ``Block turbo code performances for long-haul
  {DWDM} optical transmission systems,'' in \emph{Proc. Optical Fiber
  Communication Conference (OFC)}, Baltimore, MD, Mar. 2000.

\bibitem{Mizuochi04}
T.~Mizuochi, Y.~Miyata, T.~Kobayashi, K.~Ouchi, K.~Kuno, K.~Kubo, K.~Shimizu,
  H.~Tagami, H.~Yoshida, H.~Fujita, M.~Akita, and K.~Motoshima, ``Forward error
  correction based on block turbo code with 3-bit soft decision for 10-{G}b/s
  optical communication systems,'' \emph{IEEE J. Quantum Electron.}, vol.~10,
  no.~2, pp. 376--386, Mar./Apr. 2004.

\bibitem{Vasic02}
B.~Vasic and I.~B. Djordjevic, ``Low-density parity check codes for long-haul
  optical communication systems,'' \emph{IEEE Photon. Technol. Lett.}, vol.~14,
  no.~8, pp. 1208--1210, Aug. 2002.

\bibitem{Vasic03}
B.~Vasic, I.~B. Djordjevic, and R.~K. Kostuk, ``Low-density parity check codes
  and iterative decoding for long-haul optical communication systems,''
  \emph{J. Lightw. Technol.}, vol.~21, no.~2, pp. 438--446, Feb. 2003.

\bibitem{Djordjevic03}
I.~B. Djordjevic and B.~Vasic, ``Projective geometry {LDPC} codes for
  ultralong-haul {WDM} high-speed transmission,'' \emph{IEEE Photon. Technol.
  Lett.}, vol.~15, no.~5, pp. 784--786, May 2003.

\bibitem{ITU-T_G.975.1}
ITU, ``Forward error correction for high bit-rate {DWDM} submarine systems,''
  ITU-T Recommendation G.975.1, Tech. Rep., Feb. 2004.

\bibitem{Djordjevic09}
I.~B. Djordjevic, M.~Arabaci, and L.~L. Minkov, ``Next generation {FEC} for
  high-capacity communication in optical transport networks,'' \emph{J. Lightw.
  Technol.}, vol.~27, no.~16, pp. 3518--3530, Aug. 2009, (Invited Paper).

\bibitem{Chang10}
F.~Chang, K.~Onohara, and T.~Mizuochi, ``Forward error correction for 100 {G}
  transport networks,'' \emph{IEEE Commun. Mag.}, vol.~10, no.~3, pp. S48--S55,
  Mar. 2010.

\bibitem{Beppu15}
S.~Beppu, K.~Kasai, M.~Yoshida, and M.~Nakazawa, ``{2048 QAM (66 Gbit/s)}
  single-carrier coherent optical transmission over 150 km with a potential
  {SE} of {15.3 bit/s/Hz},'' \emph{Opt. Express}, vol.~23, no.~4, pp.
  4960--4969, Feb. 2015.

\bibitem{Qian13ofc}
D.~Qian, E.~Ip, M.-F. Huang, M.-J. Li, and T.~Wang, ``{698.5-Gb/s PDM-2048QAM}
  transmission over 3km multicore fiber,'' in \emph{Proc. Optical Fiber
  Communication Conference (OFC)}, Anaheim, CA, Mar. 2013.

\bibitem{Brueninghaus05}
K.~Brueninghaus, D.~Ast\'{e}ly, T.~S\"{a}lzer, S.~Visuri, A.~Alexiou,
  S.~Karger, and G.-A. Seraji, ``Link performance models for system level
  simulations of broadband radio access systems,'' in \emph{IEEE International
  Symposium on Personal, Indoor and Mobile Communications (PIMRC)}, Berlin,
  Germany, Sep. 2006.

\bibitem{Wan06}
L.~Wan, S.~Tsai, and M.~Almgren, ``A fading-insensitive performance metric for
  a unified link quality model,'' in \emph{IEEE Wireless Communications and
  Networking Conference (WCNC)}, Las Vegas, NV, Apr. 2006.

\bibitem{Franceschini06}
M.~Franceschini, G.~Ferrari, and R.~Raheli, ``Does the performance of {LDPC}
  codes depend on the channel?'' \emph{IEEE Trans. Commun.}, vol.~54, no.~12,
  pp. 2129--2132, Dec. 2006.

\bibitem{Leven11}
A.~Leven, F.~Vacondio, L.~Schmalen, S.~ten Brink, and W.~Idler, ``Estimation of
  soft {FEC} performance in optical transmission experiments,'' \emph{IEEE
  Photon. Technol. Lett.}, vol.~23, no.~20, pp. 1547--1549, Oct. 2011.

\bibitem{Alvarado2015_JLT}
A.~Alvarado and E.~Agrell, ``Four-dimensional coded modulation with bit-wise
  decoders for future optical communications.'' \emph{J. Lightw. Technol.},
  2015, (to appear).

\bibitem{Alvarado15a}
A.~Alvarado, E.~Agrell, D.~Lavery, and P.~Bayvel, ``{LDPC} codes for optical
  channels: Is the {``FEC Limit''} a good predictor of {Post-FEC BER}?'' in
  \emph{Proc. Optical Fiber Communication Conference (OFC)}, Los Angeles, CA,
  Mar. 2015.

\bibitem{Ryan09_Book}
W.~E. Ryan and S.~Lin, \emph{Channel Codes: Classical and Modern}.\hskip 1em
  plus 0.5em minus 0.4em\relax Cambridge University Press, 2009.

\bibitem{Jones03}
C.~Jones, A.~Matache, T.~Tian, J.~Villasenor, and R.~Wesel, ``The universality
  of {LDPC} codes on wireless channels,'' in \emph{Military Communications
  Conference}, Monterey, CA, 2003, pp. 440--445.

\bibitem{Sason09}
I.~Sason, ``On universal properties of capacity-approaching {LDPC} code
  ensembles,'' \emph{IEEE Trans. Inf. Theory}, vol.~55, no.~7, pp. 2956--2990,
  July 2009.

\bibitem{Sason11}
I.~Sason and B.~Shuval, ``On universal {LDPC} code ensembles over memoryless
  symmetric channels,'' \emph{IEEE Trans. Inf. Theory}, vol.~57, no.~8, pp.
  5182--5202, Aug. 2011.

\bibitem{Kudekar13}
S.~Kudekar, T.~Richardson, and R.~L. Urbanke, ``Spatially coupled ensembles
  universally achieve capacity under belief propagation,'' \emph{IEEE Trans.
  Inf. Theory}, vol.~59, no.~12, pp. 7761--7813, Dec. 2013.

\bibitem{Ip07}
E.~Ip and J.~M. Kahn, ``Feedforward carrier recovery for coherent optical
  communications,'' \emph{J. Lightw. Technol.}, vol.~25, no.~9, pp. 2675--2692,
  2007.

\bibitem{Viterbi98}
A.~J. Viterbi, ``An intuitive justification and a simplified implementation of
  the {MAP} decoder for convolutional codes,'' \emph{IEEE J. Sel. Areas
  Commun.}, vol.~16, no.~2, pp. 260--264, Feb. 1998.

\bibitem{Ivanov13a}
M.~Ivanov, F.~Br\"{a}nnstr\"{o}m, A.~Alvarado, and E.~Agrell, ``On the exact
  {BER} of bit-wise demodulators for one-dimensional constellations,''
  \emph{IEEE Trans. Commun.}, vol.~61, no.~4, pp. 1450--1459, Apr. 2013.

\bibitem{acikel99}
O.~A{\c c}ikel and W.~Ryan, ``Punctured turbo-codes for {BPSK}/{QPSK}
  channels,'' \emph{IEEE Trans. Commun.}, vol.~47, no.~9, pp. 1325--1323, Sep.
  1999.

\bibitem{kousa02}
M.~A. Kousa and A.~H. Mugaibel, ``Puncturing effects on turbo codes,''
  \emph{Proc. {IEE}}, vol. 149, no.~3, pp. 132--138, June 2002.

\bibitem{Vogt00}
J.~Vogt and A.~Finger, ``Improving the {max-log-MAP} turbo decoder,''
  \emph{IEEE Electronic Letters}, vol.~36, no.~23, pp. 1937--1939, Nov. 2000.

\bibitem{ETSI_EN_302_307_v121}
ETSI, ``Digital video broadcasting ({DVB}); {S}econd generation framing
  structure, channel coding and modulation systems for broadcasting,
  interactive services, news gathering and other broadband satellite
  applications ({DVB-S2}),'' ETSI, Tech. Rep. ETSI EN 302 307 V1.2.1 (2009-08),
  Aug. 2009.

\bibitem{Martinez06}
A.~Martinez, A.~{Guill\'en i F\`abregas}, and G.~Caire, ``Error probability
  analysis of bit-interleaved coded modulation,'' \emph{IEEE Trans. Inf.
  Theory}, vol.~52, no.~1, pp. 262--271, Jan. 2006.

\bibitem{Zhang14}
L.~M. Zhang and F.~R. Kschischang, ``Staircase codes with 6\% to 33\%
  overhead,'' \emph{J. Lightw. Technol.}, vol.~32, no.~10, pp. 1999--2001, May
  2014.

\bibitem{verdu94}
S.~Verd\'{u} and T.~S. Han, ``A general formula for channel capacity,''
  \emph{IEEE Trans. Inf. Theory}, vol.~40, no.~4, pp. 1147--1157, July 1994.

\bibitem{Essiambre10}
R.-J. Essiambre, G.~Kramer, P.~J. Winzer, G.~J. Foschini, and B.~Goebel,
  ``Capacity limits of optical fiber networks,'' \emph{J. Lightw. Technol.},
  vol.~28, no.~4, pp. 662--701, Feb. 2010.

\bibitem{Martinez09}
A.~Martinez, A.~{Guill\'en i F\`abregas}, G.~Caire, and F.~M.~J. Willems,
  ``Bit-interleaved coded modulation revisited: A mismatched decoding
  perspective,'' \emph{IEEE Trans. Inf. Theory}, vol.~55, no.~6, pp.
  2756--2765, June 2009.

\bibitem{Agrell10b}
E.~Agrell and A.~Alvarado, ``Optimal alphabets and binary labelings for {BICM}
  at low {SNR},'' \emph{IEEE Trans. Inf. Theory}, vol.~57, no.~10, pp.
  6650--6672, Oct. 2011.

\bibitem{Alvarado12b}
A.~Alvarado, F.~Br\"{a}nnstr\"{o}m, E.~Agrell, and T.~Koch, ``High-{SNR}
  asymptotics of mutual information for discrete constellations with
  applications to {BICM},'' \emph{IEEE Trans. Inf. Theory}, vol.~60, no.~2, pp.
  1061--1076, Feb. 2014.

\bibitem{Alvarado11b}
A.~Alvarado, F.~Br\"{a}nnstr\"{o}m, and E.~Agrell, ``High {SNR} bounds for the
  {BICM} capacity,'' in \emph{IEEE Information Theory Workshop (ITW)}, Paraty,
  Brazil, Oct. 2011.

\bibitem{Peng12_Thesis}
L.~Peng, ``Fundamentals of bit-interleaved coded modulation and reliable source
  transmission,'' Ph.D. dissertation, University of Cambridge, Cambridge, UK,
  Dec. 2012.

\bibitem{Bocherer14}
G.~B\"{o}cherer, ``Probabilistic signal shaping for bit-metric decoding,'' in
  \emph{IEEE International Symposium on Information Theory (ISIT)}, Honolulu,
  HI, July 2014.

\bibitem{Jalden10}
J.~Jald\'en, P.~Fertl, G., and Matz, ``On the generalized mutual information of
  {BICM} systems with approximate demodulation,'' in \emph{IEEE Information
  Theory Workshop (ITW)}, Cairo, Egypt, Jan. 2010.

\bibitem{Nguyen11}
T.~Nguyen and L.~Lampe, ``Bit-interleaved coded modulation with mismatched
  decoding metrics,'' \emph{IEEE Trans. Commun.}, vol.~59, no.~2, pp. 437--447,
  Feb. 2011.

\bibitem{Szczecinski12a}
L.~Szczecinski, ``Correction of mismatched {L}-values in {BICM} receivers,''
  \emph{IEEE Trans. Commun.}, vol.~60, no.~11, pp. 3198--3208, Nov. 2012.

\bibitem{Smith10}
B.~P. Smith and F.~R. Kschischang, ``Future prospects for {FEC} in fiber-optic
  communications,'' \emph{IEEE J. Quantum Electron.}, vol.~16, no.~5, pp.
  1245--1257, Sep./Oct. 2010.

\bibitem{Ivanov14_Arxiv}
M.~Ivanov, C.~H\"{a}ger, F.~Br\"{a}nnstr\"{o}m, A.~{Graell i Amat},
  A.~Alvarado, and E.~Agrell, ``On the information loss of the max-log
  approximation in {BICM} systems,'' Aug. 2014, available at
  http://arxiv.org/abs/1408.2214.

\bibitem{Menyuk87}
C.~R. Menyuk, ``Nonlinear pulse propagation in birefringent optical fibers,''
  \emph{IEEE J. Quantum Electron.}, vol.~23, no.~2, pp. 174--176, Feb. 1987.

\bibitem{Maher15}
R.~Maher, T.~Xu, L.~Galdino, M.~Sato, A.~Alvarado, K.~Shi, S.~J. Savory, B.~C.
  Thomsen, R.~I. Killey, and P.~Bayvel, ``Spectrally shaped {DP-16QAM}
  super-channel transmission with multi-channel digital back-propagation,''
  \emph{Sci. Rep.}, vol.~5, pp. 1--8, Feb. 2015, article number: 8214.

\end{thebibliography}
\bibliographystyle{IEEEtran}

\end{document}